\documentclass[journal=jctc,manuscript=article]{achemso}

\usepackage{booktabs}
\usepackage[version=3]{mhchem} 
\usepackage{xcolor}
\usepackage{amsmath,amssymb}
\usepackage{comment}
\usepackage{multirow}
\usepackage{tablefootnote}
\usepackage{ulem}

\usepackage{subcaption} 

\SectionNumbersOn

\author{Mike Devereux}
  \altaffiliation{These authors contributed equally}

\author{Eric D. Boittier}
  \altaffiliation{These authors contributed equally}

  \author{Markus Meuwly}
  \altaffiliation{Department of Chemistry, Brown
  University, USA}
\affiliation[University of Basel]{Department of Chemistry, University
  of Basel, Klingelbergstrasse 80, CH-4056 Basel, Switzerland.}
\email{m.meuwly@unibas.ch}

\title{Systematic Improvement of Empirical Energy Functions in the Era
  of Machine Learning}


\begin{document}

\bibliographystyle{apsrev}

\begin{abstract}
The impact of targeted replacement of individual terms in conventional
empirical force fields is quantitatively assessed for pure water,
solvated K$^+$ and Cl$^-$ ions, and dichloromethane. For the
electrostatics, point charges (PCs) and machine learning-based
minimally distributed charges (MDCM) fitted to the molecular
electrostatic potential are evaluated together with electrostatics
based on the Coulomb integral. The impact of explicitly including
second-order terms is investigated by adding a fragment molecular
orbital (FMO)-derived polarization energy to an existing force field,
in this case CHARMM. It is demonstrated that anisotropic
electrostatics reduce the RMSE for water (by 1.6 kcal/mol),
dichlormethane (by 0.8 kcal/mol) and for solvated Cl$^-$ clusters (by
0.4 kcal/mol). An additional polarization term can be neglected for
dichloromethane but notably improves errors in pure water (by 1.1
kcal/mol) and in Cl$^-$ clusters (by 0.4 kcal/mol) and is key to
describing solvated K$^+$, reducing the RMSE by 2.3 kcal/mol. A 12-6
Lennard-Jones functional form is found to work satisfactorily with PC
and MDCM electrostatics, but is not appropriate for descriptions that
account for the electrostatic penetration energy. The importance of
many-body contributions is assessed by comparing a strictly 2-body
approach with self-consistent reference data. Dichloromethane can be
approximated well with a 2-body potential while water and solvated
K$^+$ and Cl$^-$ systems require explicit many-body
corrections. Finally, a many-body-corrected dimer PES is found to
exceed the accuracy attained using a conventional empirical force
field, potentially reaching that of an FMO calculation. The present
work systematically quantifies which terms improve the performance of
an existing force field and what reference data to use for
parametrizing these terms in a tractable fashion for machine-learned
(ML) fitting of pure and heterogeneous systems.
\end{abstract}

\section{Introduction}
Empirical force fields owe their success to striking a delicate
balance between acceptable accuracy in modelling the desired processes
while remaining sufficiently simple to apply to large systems and long
timescales. As empirical FFs do not explicitly include electrons it is
possible to use them for simulations which are out of reach for
quantum-based methods, but continued development of computer hardware
has tended to outpace force field development. While existing force
fields are exploiting new hardware to simulate ever larger systems and
longer timescales,\cite{DEShawTrypsin} increased computational power
also provides an opportunity to add more detail to refine force fields
that are still applicable to systems and timescales of chemical or
biochemical interest. Crucial to this task is determining to what
extent increasing the detail of each term provides greater accuracy in
describing the intermolecular interactions. At the same time, this
description needs to be widely applicable and possible to parametrize
for even highly heterogeneous chemical systems, and remain
computationally tractable for a broad range of timescales. Much
ongoing work is currently devoted to this
task.\cite{ponder:2010,physnet,paesani:2013,behler:2007,sibfaMD,ANI-2x}\\

\noindent
One rather underexplored possibility is to replace existing terms in
empirical force fields by more rigorous approximations. If such an
approach yields a substantial improvement in the quality of the energy
function it is useful to invest effort to improve that term. This
strategy is pursued and quantitatively tested in the present work for
the CHARMM empirical energy function.\cite{charmm,Brooks:2009} As well
as improving existing terms the impact of adding missing polarization
effects is examined using fragment molecular orbital (FMO)
embedding.\\

\noindent
Successful parametrization of (empirical) energy functions is often
judged by either comparing reference properties from electronic
structure calculations with those determined from the fitted model, or
by its ability to approximately describe experimental observables
(e.g. diffusion coefficients or infrared spectra), or in combination.
Developing and maintaining empirical force fields remains a tremendous
task, involving large and diverse research groups. To improve such
empirical energy functions, increasing numbers of atom types need to
be introduced and corresponding atom-specific parameters are
required. On the other hand, machine learning-based techniques gain
traction and provide accurate and versatile
models.\cite{physnet,paesani:2013,behler:2007,ANI-2x} The accuracy of
these models, once trained, primarily relies on the available
reference data. Another potentially promising route is to start with
physics-based representations of intermolecular interactions that
depend on models and parameters. In a second step, machine learning
can be used to further refine the models. One open question for
ML-based models concerns the capabilities to extrapolate and
generalize such approaches to chemical systems they were not trained
for. For conventional empirical force fields, this is usually
addressed by introducing new atom types. One potential solution,
starting from a 2-body description of the PES and adding a many-body
polarization correction,\cite{paesani:2013,paesani2022,Paesani:2023}
is examined here. \\

\noindent
In addition, the present work follows a middle way in that each term
of an empirical energy function is queried as to its sensitivity with
respect to reference data from electronic structure calculations at a
given level of theory. If replacing such a term yields an improved
model from comparing with the underlying reference data, the
reparametrization effort is deemed worthwhile. As with ML-based
approaches, employing a higher level of electronic structure theory
primarily affects computational effort at the level of reference data
generation. Hence, a sufficiently robust level of theory, applicable
to molecular clusters of ca. 20 monomers is chosen. Energy functions
fitted to reference data within ``chemical accuracy" ($\sim 1$
kcal/mol for energies) allow running atomistic simulations at the
level of the reference data, but at speeds comparable to evaluation
times of an empirical energy function. This has also been shown for
small molecules for which now CCSD(T)-quality MD simulations on the
microsecond time scale have become possible.\cite{kaeser.mm:2023} \\

\noindent
First, the methods and concepts employed are presented. This is
followed by results quantifying the effect of substituting individual
terms in the empirical force field expression to obtain increasingly
accurate inter- and intramolecular interactions. Next, including
many-body effects using FMO is explored. Finally, the results are
discussed in the broader perspective of modelling energy functions in
the machine learning era.\\

\section{Methods}
In the following section the individual contributions considered are
first defined. These correspond to terms in widely used empirical
force fields. An additional FMO-based definition of the polarization
energy is then described, and subsequently extended as a many-body
correction to a dimer potential energy surface. Next, the simulations
used to obtain H$_{2}$O, CH$_{2}$Cl$_{2}$, microsolvated K$^+$ and
Cl$^-$ test systems are detailed. Finally, the fitting procedure to
adjust Lennard-Jones (LJ) or double-exponential
(DE)\cite{Wu_Brooks_2019} parameters is summarized.\\

\subsection{Reference Bonded and Nonbonded Energies}
\label{sec:bonded}
Empirical force fields often separate the total (condensed-phase)
system energy into internal (bonded) and external (nonbonded)
energies.\cite{case:2005,MacKerell:2010,christen:2005} Internal
energies of monomers may contain, among others, contributions due to
harmonic bonded, angle, dihedral, improper, Urey-Bradley and
intramolecular electrostatic and van der Waals (vdW) terms between
atoms separated by at least three bonds. These monomer energies
usually do not include or explicitly couple to the `nonbonded'
interactions with surrounding moieties. Hence, their contribution
describes only the response of the isolated monomer to changes in
geometry.\\

\noindent
In the present work corresponding analogs to FF bonded terms were
determined from electronic structure calculations by evaluating the
change in energy between a geometry-optimized gas-phase monomer
$V_{\rm mon,ref}$ and the energy $V_{{\rm mon},i}$ of each of the
$N_{\rm mol}$ monomer conformers $i$ extracted from the system of
interest:
\begin{equation}
V_{\rm bond,qm} = \sum_{i=1}^{N_{\rm mol}} (V_{{\rm mon},i} - V_{\rm
  mon,ref}) \neq V_{\rm bond,ff}
\end{equation}
The quantity $V_{\rm bond,qm}$ from electronic structure calculations
is not yet directly comparable with force field bonded energies
$V_{\rm bond,ff}$ as these values may also contain intramolecular
electrostatic and van der Waals contributions between atoms separated
by more than 2 bonds. However, after subtracting the mean of each
distribution the two sets of energies can be compared:
\begin{equation}
V_{\rm bond,ff} - \bar{V}_{\rm bond,ff} \approx V_{\rm bond,qm} -
\bar{V}_{\rm bond,qm}
\label{Eq:Ebond}
\end{equation}
Here, $\bar{V}_{\rm bond}$ is the mean bonded energy of the set to be
compared. The mean is preferred as a reference over the minimum bonded
energy of the set as this structure may be an outlier that is poorly
described by the force field relative to electronic structure
theory.\\

\noindent
Similarly, subtracting the bonded (isolated monomer energy)
contributions from the total electronic structure energy yields
the intermolecular (nonbonded) interaction energy, which can be
used to fit nonbonded force field terms and to quantify the residual
`nonbonded error' with the bonded term removed from the FF:
\begin{equation}
V_{\rm nbond,qm} = V_{\rm qm} - \sum_{i=1}^{N_{\rm mol}}{V_{{\rm
      mon},i}}
\label{Eq:Nbond}
\end{equation}
For force fields the nonbonded terms can be evaluated by omitting the
bonded terms from the energy function, for example:
\begin{equation}
V_{\rm nbond,ff} = \sum_{I=1}^{N_{\rm mol}}\sum_{J>I}^{N_{\rm mol}}
\sum_{i=1}^{N_{{\rm atm},I}}\sum_{j=1}^{N_{{\rm atm},J}}{(V_{{\rm
      vdW,ff,}i,j} + V_{{\rm elec,ff,}i,j})}
\label{Eq:chmNbond}
\end{equation}
Here, the sum over $I,J$ includes all pairs of monomers in the system,
whereas the sum $i,j$ runs over the vdW and electrostatic terms
$V_{\rm vdW,ff}$ and $V_{\rm elec,ff}$ of all $N_{\rm atm}$ atoms of
monomers $I$ and $J$.\\

\subsection{Reference Electrostatic Terms}
The electrostatic interaction term in empirical force fields such as
CHARMM\cite{charmm,MacKerell:2010}, GROMACS\cite{van:2005} or
AMBER\cite{case:2005} has commonly been described using point charges
at nuclear positions. The charges do not usually respond to changes in
molecular conformation or to changes in the external electric field
created by surrounding molecules (i.e. such force fields are not
polarizable), although exceptions to this
exist.\cite{Jensen:2022,yu:2013,MM.fmdcm:2022} As such, electrostatic
energies from atom-centered charges are conceptually related to a sum
of interactions between frozen monomer charge densities:
\begin{align}
E_{\rm elec} =& \sum_{I=1}^{N_{\rm mol}}\sum_{J>I}^{N_{\rm mol}}
\ \Bigl[ \iint\displaylimits_{\Omega_I \Omega_J}\frac{\rho_{0,I}({{\bf
        r}_{I}})\rho_{0,J}({\bf r}_{J})}{r_{IJ}} d{{\bf r}_{I}}d{{\bf
      r}_{J}} + \sum_{j=1}^{N_{{\rm atm},J}}
  \int\displaylimits_{\Omega_I} \frac{\rho_{0,I}({\bf r}_{I})
    Z_{j}}{r_{Ij}}d{{\bf r}_{I}} \nonumber \\ &+ \sum_{i=1}^{N_{{\rm
        atm},I}} \int\displaylimits_{\Omega_J} \frac{\rho_{0,J}({\bf
      r}_{J}) Z_{i}}{r_{iJ}}d{\bf r}_{J} + \sum_{i=1}^{N_{{\rm
        atm},I}} \sum_{j=1}^{N_{{\rm atm},J}} \frac{Z_{i}
    Z_{j}}{r_{ij}} \Bigr]
\label{Eq:Eel} \\
E_{\rm elec} \approx & \sum_{I=1}^{N_{\rm mol}}\sum_{J>I}^{N_{\rm
    mol}}\sum_{i=1}^{{N_{{\rm atm},I}}}\sum_{j=1}^{N_{{\rm
      atm},J}}\frac{q_{i}q_{j}}{r_{ij}}
\label{Eq:Coulomb}
\end{align}
where here and in the following $q_{i}$ is the point charge associated
with atom $i$, $r_{ij}$ is the distance between atoms $i$ and $j$,
$r_{Ij}$ is the distance between point ${\bf r}_{I}$ in the molecular
volume of monomer $I$ and nucleus $j$ with charge $Z_{j}$, $\Omega_I$
is the molecular volume of monomer $I$, $\rho_{0,I}({\bf r}_I)$ is the
electron density of frozen monomer $I$ at point ${\bf r}_{I}$ inside
that volume.\\

\noindent
These frozen monomer densities are typically determined for gas-phase,
unpolarized molecules. In empirical force fields a degree of averaged
`prepolarization' can be introduced to better represent electrostatics
in condensed phases,\cite{horn:2004,jorgensenTIPS:1981,berendsen:1987}
but only at the cost of decreased accuracy in environments that do not
resemble the condensed phase.\cite{morozenko:2014,xantheasPCCP:2023}
Here, gas-phase densities are employed to provide reference data for
fitting atom-centered point charges and `minimally distributed charge
models'
(MDCMs)\cite{MM.dcm:2014,MM.mdcm:2017,MM.MDCM:2020,MDCM-Git,MM.fmdcm:2022}
that - similar to multipole
moments\cite{stone:2005,MM.mtp:2012,kramer:2013,MM.mtp:2013,cardamone:2014,SIBFA2}
- more accurately approximate $E_{\rm elec}$ than atom-centered point
charges alone as they better describe anisotropic electrostatic
interactions. MDCM uses differential evolution, an energy-based global
optimization machine learning technique, to determine the magnitude
and position of a user-specified number of off-center point charges to
best reproduce the ESP.\\

\noindent
In addition to fitting anisotropic charge models, reference $E_{\rm
  elec}$ values were evaluated directly from the Coulomb integral for
each pair of monomers using an adapted version of the `Orbkit'
program\cite{orbkit} interfaced with libcint\cite{libcint}, to replace
$V_{\rm elec,ff}$ in eq. \ref{Eq:chmNbond}. As such the accuracy of a
force field including different approximate electrostatic interaction
terms can be compared with one including exact Coulomb integral
reference data from electronic structure theory.\\

\noindent
It should be noted that so-called `penetration' effects are
encountered in the Coulomb integral at close range due to reduced
shielding of nuclear charge as monomer densities overlap. This is not
accounted for in the point charge or distributed charge models used
here, but could be added as a correction if necessary, as in the
`SIBFA' energy function.\cite{SIBFA3}\\

\subsection{Reference Polarization}
\label{sec:polarization}
Numerous approaches to describe the response of a monomer electron
density to both external electric fields generated from molecular
environments\cite{yu:2013,Thole1981} and due to changes in monomer
conformation\cite{Patel:2004,Rick:1994,Jensen:2022,Chengwen:2020,MM.fmdcm:2022}
exist. Here, a polarization correction based on FMO electrostatic
embedding is used that accounts for changes in electrostatic
interaction due to distortions in the electron cloud of each monomer
and also resulting destabilization of the polarized monomer
energies. This is expected to provide a more complete description than
typical approximations\cite{Thole1981} based on induced dipole moments
only.\\

\begin{figure}
    \centering
    \includegraphics[width=0.8\textwidth]{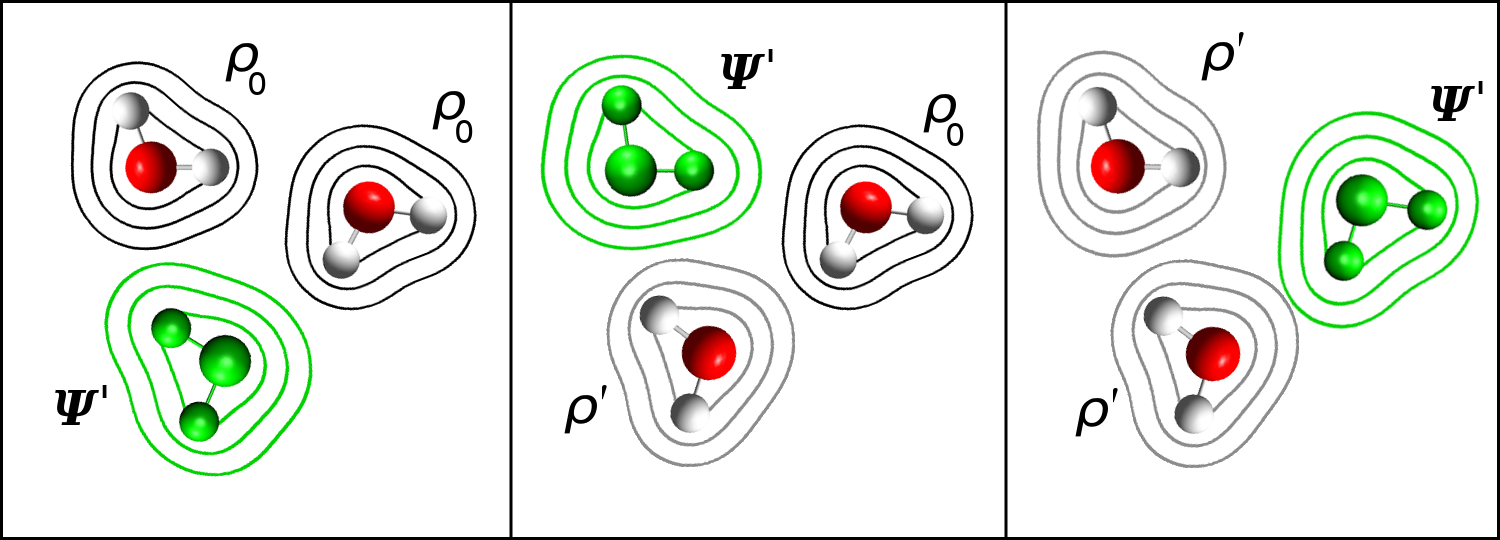}
    \caption{The first cycle of the FMO embedding process. Each monomer
      wave function $\Psi'$ is re-evaluated in the presence of
      gas-phase densities $\rho_{0}$ of the surrounding monomers. Once
      a monomer wave function has been updated the perturbed density
      $\rho'$ is used to provide a background field for the remaining
      monomers. The process is repeated until monomer wave functions
      no longer change significantly.}
    \label{fig:FMO_schema}
\end{figure}

\noindent
While various energy decomposition approaches offer alternative
definitions for the polarization
energy,\cite{SAPT,gordon:1996,morokuma:1976} FMO
embedding\cite{FMO:1999} is well-suited to defining the response term
that is missing from electrostatic interactions between frozen
gas-phase monomers. Embedding begins by reevaluating monomer wave
functions $(\Psi')$ in the presence of the frozen electron densities
$\rho_0$ of surrounding monomers to obtain new `embedded' monomer
densities $\rho'$, see Figure \ref{fig:FMO_schema}. The new densities
provide an updated background electric field to obtain the next set of
monomer electron densities and this cycle continues until embedded
monomer densities and the electric field of their neighbors become
self-consistent. No orthogonalization or antisymmetrization takes
place between distinct monomers during embedding as the monomers
interact only with their electric field. Similar approaches have been
shown to perform well for water cluster energies in electronic
structure calculations.\cite{truhlarEmbedding}\\

\noindent
The electronic Hamiltonian describing each polarized (embedded)
monomer $I$ can be written:
\begin{equation}
H_{I} = \sum_{i}^{n_{I}} {\Bigl\{ -\frac{1}{2} \nabla^{2}_{i} -
  \sum_{s}^{N_{\rm atm}} \frac{Z_{s}}{{|{{\rm \bf r}_{i}} - {{\rm \bf
          r}_{s}}|}} + \sum_{J \neq I}^{N_{\rm mol}}
  \int\frac{\rho''_{J}({\rm\bf r'})}{{|{{\rm \bf r}_{i}} - {{\rm \bf
          r'}}|}}d{\rm \bf r'} \Bigr\}} +
\sum_{i>j}^{n_{I}}\frac{1}{{|{{\rm \bf r}_{i}} - {{\rm \bf r}_{j}}|}}
\label{Eq:fmo_hamiltonian}
\end{equation}
where $i$ runs over all of the $n_{I}$ electrons of monomer $I$. The
kinetic energy operator $\Bigl\{ -\frac{1}{2} \nabla^{2}_{i}$ acts on
electron $i$, $s$ runs over all nuclei of all monomers in the system
(not just $I$) to describe the interaction between electron $i$ and
each nuclear charge $Z_{s}$ at position ${\rm \bf r}_{s}$, and $J$
runs over integrals of the embedded electron densities $\rho''_{J}$ of
each monomer with electron $i$. The pairwise sum $i>j$ accounts for
electron-electron repulsion within monomer $I$. The approach is
implemented in GAMESS\cite{GAMESS} with DFT support.\\

\noindent
Subtracting gas-phase monomer energies from these embedded monomer
energies yields an electrostatic + polarization energy. This contains
the change in internal monomer energies and the Coulomb integral of
the charge density of each polarized monomer with each of the other
polarized monomers. The Coulomb integral (evaluated using Orbkit with
libcint) between the gas-phase (frozen) monomer densities can then be
subtracted to yield a polarization energy. This is
the change in the Coulomb integral after embedding each monomer in the
electric field of its neighbors, plus the change in monomer internal
energies:
\begin{align}
E_{\rm elec + pol}^{\rm FMO} &= \sum_{I=1}^{N_{\rm
    mol}}\sum_{I>J}^{N_{\rm mol}} \Biggl[
  \ \iint\displaylimits_{\Omega_{I},\Omega_{J}} \frac{\rho''_{I}({\rm
      \bf r}_{I}) \rho''_{J}({\rm \bf r}_{J})}{r_{IJ}} dr_{I}dr_{J} +
  \sum_{j=1}^{N_{\rm atm},J} \int\displaylimits_{\Omega_I}
  \frac{\rho''_{I}({\bf r}_{I}) Z_{J}}{r_{Ij}}d{{\bf r}_{I}} +
  \nonumber \\ &\sum_{i=1}^{N_{\rm atm},I}
  \int\displaylimits_{\Omega_J} \frac{\rho''_{J}({\bf r}_{J})
    Z_{i}}{r_{iJ}}d{{\bf r}_{J}} + \sum_{i=1}^{N_{\rm atm},I}
  \sum_{j=1}^{N_{\rm atm},J} \frac{Z_{i} Z_{j}}{r_{ij}} \Biggr] +
\sum_{I=1}^{N_{\rm mol}} \Biggl[ \ \tilde{E}_{I} - E_{0,I} \Biggr]
\\ E_{\rm pol}^{\rm FMO} &= E_{\rm elec + pol}^{\rm FMO} - E_{\rm
  elec}
\label{Eq:Epol_FMO}
\end{align}
where $\tilde{E}_{I}$ is the energy of an embedded monomer according
to Eq. \ref{Eq:fmo_hamiltonian} minus the Coulomb interaction with the
electric field of the other monomers. Energy $E_{0,I}$ is the
gas-phase (non-polarized) monomer energy and $E_{\rm elec}$ was
defined in eq. \ref{Eq:Eel}.\\

\noindent
The second-order polarization energy $E_{\rm pol}^{\rm FMO}$ can be
added to the first-order energy terms of empirical force fields,
according to:
\begin{equation}
V_{\rm tot} = V_{\rm bond,ff} + V_{\rm elec,ff} + V_{\rm vdW,ff} +
E_{\rm pol}^{\rm FMO}
\end{equation}
In this way a residual error in the force field PES can be computed
once the polarization response to the electric field of all
surrounding moieties is explicitly accounted for.\\

\subsection{Many-body-corrected 2-body potentials}
FMO will also be used to correct a 2-body interaction by
adding missing many-body contributions. The aim is to compare the
maximum accuracy that can be obtained using a
many-body-corrected 2-body approach with the maximum accuracy
possible using a typical force field functional form, with or without
a polarization correction.\\

\noindent
The 2-body interaction energy over all dimers in a cluster
is
\begin{equation}
E_{\rm int}^{\rm 2body} = \sum_{I=1}^{N_{\rm mol}}\sum_{I>J}^{N_{\rm
    mol}} [E_{IJ}^{\rm dimer} - E_{0,I} - E_{0,J}]
\label{Eq:2body}
\end{equation}
Here, $E^{\rm dimer}$ is the dimer energy from which the monomer
energies $E_{0,I}$ and $E_{0,J}$ of both gas-phase monomers is
subtracted to yield a 2-body cluster energy that is a sum of all
pairwise interactions without many-body effects.\\

\noindent
An exact many-body correction to this sum-of-dimers, or ``2-body''
approximation would again need to account for many-body Pauli
effects. As electrons are not explicitly described, chemical
atom-types and their relative locations would be required as inputs to
such a function. This approach suffers from permutational explosion in
the required reference data and is avoided here. One common
simplification is to restrict the many-body correction to only the
leading terms (truncating at 3- or occasionally 4-body
corrections).\cite{bowman:2021} Nonetheless, this adds significant
additional training effort and computational cost for a term that
typically contributes one or two orders of magnitude less to the total
energy than the 1$^{\rm st}$-order or 2-body term.\cite{bowman:2021}\\

\noindent
A term based on FMO embedding is used to transform the 2-body
polarization, which arises within each dimer, into the many-body
polarization that arises as a response to electrostatic interactions
within the whole cluster. This many-body correction is not complete as
remaining Pauli contributions to the many-body energy are neglected,
but offers a good approximation. To obtain the corrective term from
FMO calculations, the 2-body dimer polarization energy $E_{\rm
  elec+pol}^{\rm FMO,2body}$ is first removed from the 2-body cluster
energy $E_{\rm int}^{\rm 2body}$ and the many-body polarization energy
$E_{\rm elec+pol}^{\rm FMO}$ (eq. \ref{Eq:Epol_FMO}) is then added
back on. $E_{\rm elec+pol}^{\rm FMO,2body}$ and the final corrective
term $E_{\rm int}^{\rm 2body,corr}$ are defined as
\begin{align}
E_{\rm elec+pol}^{\rm FMO,2body} &= \sum_{I=1}^{N_{\rm
    mol}}\sum_{I>J}^{N_{\rm mol}} \Bigl[
  \ \iint\displaylimits_{\Omega_{I},\Omega_{J}} \frac{\rho_{{\rm dimer},I}''({\rm \bf r}_{I}) \rho_{{\rm dimer},J}''({\rm \bf
      r}_{J})}{r_{IJ}} d{\rm\bf r}_{I}d{\rm\bf r}_{J} +
  \sum_{j=1}^{N_{{\rm atm},J}}\int\displaylimits_{\Omega_{I}}
  \frac{\rho_{{\rm dimer,}I}''({\rm \bf r}_{I}) Z_{j}}{r_{Ij}}
  d{\rm\bf r}_{I} + \nonumber \\ & \sum_{i=1}^{N_{{\rm
        atm},I}}\int\displaylimits_{\Omega_{J}} \frac{\rho_{{\rm dimer,}J}''({\rm \bf r}_{J}) Z_{i}}{r_{iJ}} d{\rm\bf r}_{J} +
  \sum_{i=1}^{N_{{\rm atm},I}}\sum_{j=1}^{N_{{\rm atm},J}} \frac{Z_{i}
    Z_{j}}{r_{ij}} + \tilde{E}_{I}^{\rm dimer} + \tilde{E}_{J}^{\rm
    dimer} - E_{0,I} - E_{0,J} \Bigr] \label{Eq:epol_corr} \\
E_{\rm int}^{\rm 2body,corr} &= E_{\rm int}^{\rm 2body} - E_{\rm
  elec+pol}^{\rm FMO,2body} + E_{\rm elec+pol}^{\rm FMO}
\label{Eq:mb_pol_corr}
\end{align}
where $\rho_{\rm dimer}''$ is the electron density of one monomer
embedded in the electric field of its dimer neighbor
only. $\tilde{E}^{\rm dimer}$ is the energy of an embedded monomer in
a dimer, excluding the Coulomb interaction with the electric field of
its partner, evaluated using Orbkit, libcint and libxc\cite{libxc}
from the FMO molecular orbitals. $E_{\rm elec+pol}^{\rm FMO,2body}$
and $E_{\rm elec+pol}^{\rm FMO}$ are combined in
eq. \ref{Eq:mb_pol_corr} with the 2-body interaction energy $E_{\rm
  int}^{\rm 2body}$ to yield the final many-body corrected interaction
energy $E_{\rm int}^{\rm 2body,corr}$.\\

\subsection{Electronic Structure Calculations}
The PBE0 hybrid functional\cite{PBE0} together with the aug-cc-pVDZ
basis set was chosen to provide accurate descriptions of the electron
density, required for fitting PC and MDCM charge
models.\cite{Perdew:2017} These models were generated from reference
PBE/aug-cc-pVDZ calculations on an ensemble of monomer structures
using molecular symmetry by following published
protocols.\cite{MM.MDCM:2020,MM.fmdcm:2022} Such calculations also
remain feasible for systems large enough to be reasonably
representative of the condensed phase (i.e. including ``buried''
monomers completely surrounded by others).\\

\noindent
Monomer, dimer and cluster energies were computed using
Molpro\cite{molpro1,molpro2,molpro3}. The findings are expected to
transfer to other levels of theory. FMO calculations and energies of
monomers, clusters and dimers were also calculated using GAMESS
2022\cite{GAMESS} with basis sets defined manually to match the
definitions from Molpro (sample Molpro and GAMESS input files are
provided as supplementary information). GAMESS was used primarily for
FMO calculations, while Molpro was preferred for exporting MOs to
Orbkit.\\

\noindent
Tests with counterpoise corrected interaction energies were performed
using Gaussian 16\cite{g16}. Aside from a systematic shift, BSSE
determined for water clusters did not alter results sufficiently
(Figure \ref{fig:si_bsse}) to justify the substantial additional
computational cost. Counterpoise calculations were therefore not
performed for the ionic or CH$_{2}$Cl$_{2}$ systems, and results
reported in the main text are not counterpoise-corrected.\\

\subsection{Sampling Cluster Geometries}
Water clusters of 20 molecules were obtained from molecular dynamics
(MD) simulations of a periodic water box with side length 41 \AA~
using MDCM electrostatics.\cite{MM.mdcm:2017,MM.MDCM:2020} All MD
simulations were carried out using CHARMM with periodic boundary
conditions, with a time step of $\Delta t = 1$ fs, and particle mesh
Ewald summation. Nonbonded interactions were truncated at 16 \AA\/ and
the simulations were carried out in the $NpT$ ensemble at 300 K.
Structures were obtained by randomly selecting a molecule and saving
the coordinates of the 19 nearest neighbors. A similar procedure was
performed for dichloromethane, using PC electrostatics and bonded
parameters from CGenFF\cite{MacKerell:2010} without DRUDE sites, where
20 molecules were extracted from MD simulations of 254 molecules. For
the ionic systems, clusters of radius 4.6 \AA~ centered on individual
K$^{+}$ and Cl$^{-}$ ions were extracted from MD simulations in a
periodic water box containing 2 of each ion and 536 water molecules,
yielding clusters with between 12 and 24 water molecules surrounding a
central ion.\\

\subsection{Fitting Lennard-Jones Parameters}
A python-based energy evaluation routine, written in
Jax,\cite{jax2018github} was developed and was validated against
CHARMM interaction energies for water and dichloromethane clusters to
be within $\pm 10^{-7}$ kcal/mol (machine accuracy).  Lennard-Jones
parameters ($\sigma$, $r_{\mathrm{min}}/2$) were optimized using the
Nelder-Mead algorithm, a downhill simplex method, with a tolerance of
$10^{-6}$, as implemented in SciPy.\cite{2020SciPy-NMeth} To prevent
biased results favoring available parameter sets such as CGenFF, which
may not be appropriate for electrostatic models such as the Coulomb
integral, values of $\sigma$ were sampled randomly from a uniform
distribution between 0.25 to 2.5 \AA\/; $\epsilon$ values ranged from
0.001 to 1.0 kcal/mol. Optimizations were restricted to $10^5$
function evaluations. Each optimization was repeated 1000 times,
retaining the best-performing model for further use. The DE pair
potential was also investigated, using the same functional form
available in the Open Force Field Toolkit.\cite{DE_2023} \\

\section{Results and Discussion}
In the following section reference data are first established using
widely used force field parameters. Next the reference terms described
in sections \ref{sec:bonded} to \ref{sec:polarization} are introduced
into the CHARMM force field one by one in place of the existing terms
in order to assess the level of improvement that can be achieved with
each substitution. Finally, a 2-body potential without and with an FMO
many-body polarization term is evaluated as a possible alternative to
the empirical force field functional form.

\subsection{Force Field Reference Values}
\begin{figure}
    \centering
    \includegraphics[width=0.8\textwidth]{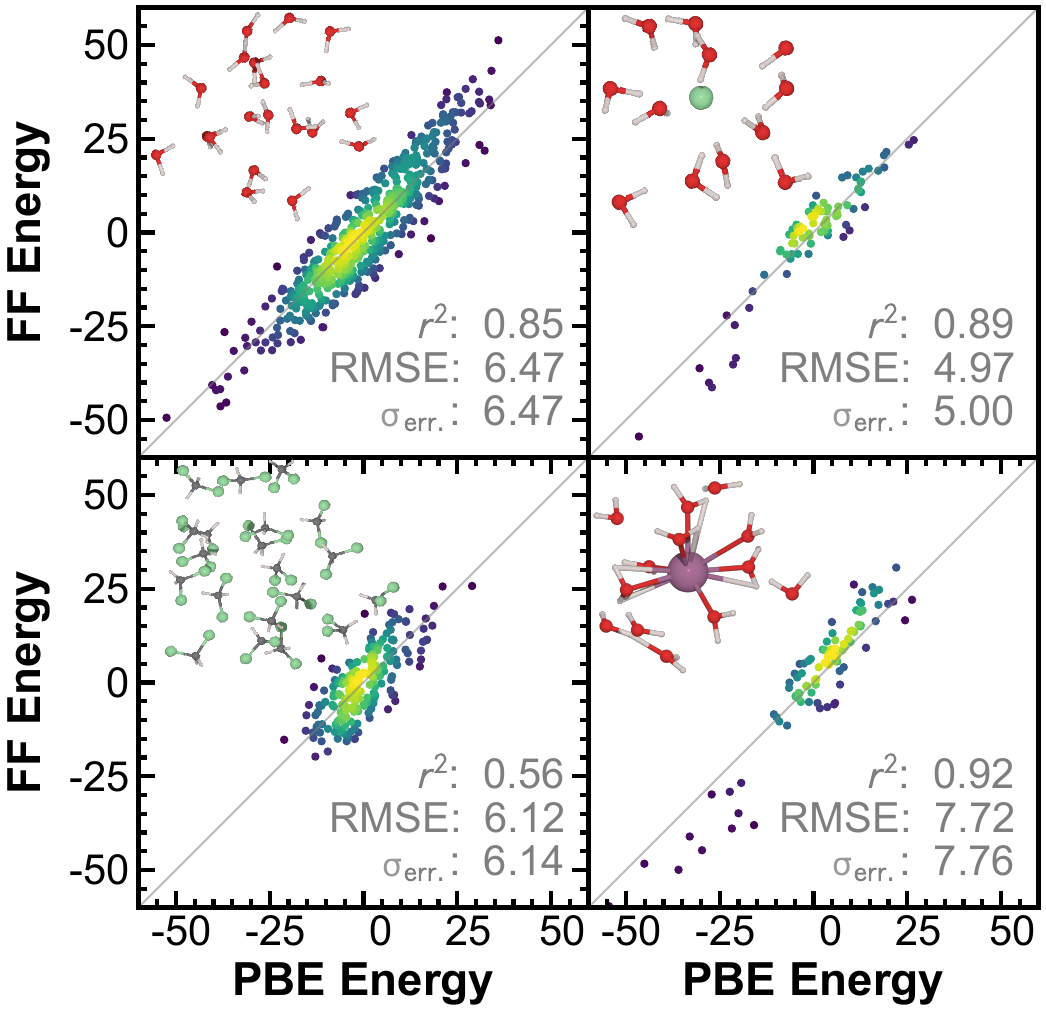}
    \caption{Comparison of total cluster energies using
      CGenFF\cite{MacKerell:2010} parameters, TIP3P for
      water\cite{jorgensenTIPS:1981} and ion parameters from Roux et
      al.\cite{beglov:1994} with PBE0 data. Results for water clusters
      (top left), solvated K$^{+}$ (top right) and Cl$^{-}$ ions
      (bottom right) and CH$_{2}$Cl$_{2}$ (bottom left) are shown,
      mean cluster energies are subtracted to allow direct
      comparison. The color code for the points reflects their density
      with yellow being regions of maximum density.}
\label{fig:cgenff_vs_dft}
\end{figure}

Figure \ref{fig:cgenff_vs_dft} compares reference PBE0 cluster
energies with those obtained using TIP3P\cite{jorgensen:1983} for
H$_{2}$O with the corresponding K$^+$ and Cl$^-$ parameters of Roux
and coworkers\cite{roux:1994} for solvated ions, and CGenFF parameters
for CH$_{2}$Cl$_{2}$ that were generated using tools from SILCS-Bio
(v. 2023.1).\cite{Vanommeslaeghe_2010, Goel_2021}\\

\noindent
As total cluster energies including bonded terms and DFT formation
energies are compared, the mean cluster energy of each distribution is
subtracted. The RMSE of 5 to 8 kcal/mol for cluster energies spanning
40 kcal/mol (CH$_{2}$Cl$_{2}$) to 80 kcal/mol (H$_{2}$O, K$^+$ and
Cl$^-$) show that although the force field parameters were not
originally fitted to these clusters or at this level of theory,
results are nonetheless meaningful with relative energies broadly
correctly described. However, each data point lying away from the
diagonal represents a cluster geometry with an energy that is over- or
underestimated relative to other geometries in the distribution. The
goal in the following sections is to determine how to maximally narrow
the width of the error distribution relative to this baseline. This is
quantified by a combination of the RMSE of the distribution (which
measures the absolute difference between a model prediction and the
reference data) and the standard deviation (width) of the error
distribution ($\sigma_{\rm error}$) which differs from the RMSE if
there is a systematic shift in the model data relative to the
reference.\\

\subsection{Substituting force field terms}
{\bf Bonded Terms:} Removing monomer energy contributions from the
reference PBE0 cluster energies and removing bonded terms from force
field results eliminates contributions to the RMSE due to
conformational changes of the cluster monomers. Table \ref{tab:comp}
and Figure \ref{fig:si_cgenff_vs_dft_nbonds} show that $\sigma_{\rm
  error}$ changes little after removing the contribution from bonded
terms, and with the exception of CH$_{2}$Cl$_{2}$ even increases
slightly. This indicates some small degree of error cancellation with
the remaining parameters. The much larger increase in RMSE, from $\sim
6$ kcal/mol to $\sim 25$ kcal/mol, compared to $\sigma_{\rm error}$
results from the fact that the nonbonded interaction energy is now
directly comparable without subtracting the mean of each
distribution. As LJ parameters are not refitted a substantial
systematic shift in the force field data relative to PBE0 is
visible. The systematic shifts will affect absolute properties such as
cluster formation energies, but not relative cluster energies.\\

\begin{table}
\resizebox{\textwidth}{!}{%
\begin{tabular}{|l|c|c|c|c|c|c|c|c|}
\hline
\multicolumn{1}{|c}{\multirow{2}{*}{Model}} & \multicolumn{2}{|c|}{H$_{2}$O} &\multicolumn{2}{|c|}{CH$_{2}$Cl$_{2}$} &\multicolumn{2}{|c|}{K$^{+}$} &\multicolumn{2}{|c|}{Cl$^{-}$}\\\cline{2-9} 
 & RMSE & $\sigma_{\rm error}$ & RMSE & $\sigma_{\rm error}$ & RMSE & $\sigma_{\rm error}$ & RMSE & $\sigma_{\rm error}$ \\
 \hline
 FF$_{\rm ref}$* & 6.5 & 6.5 & 6.1 & 6.1 & 5.0 & 5.0 & 7.7 & 7.7 \\
 \hline
 $E_{\rm bond}$+FF$_{\rm ref}$ & 27.7 & 6.6 & 23.3 & 5.7 & 22.9 & 5.4 & 25.0 & 7.9 \\
 $E_{\rm bond}$+$E_{\rm PC}$+LJ$_{\rm fit}$ & 5.8 & 5.7 & 2.1 & 2.1 & 5.0 & 5.0 & 4.2 & 4.2 \\
 $E_{\rm bond}$+$E_{\rm MDCM}$+LJ$_{\rm fit}$& 4.2 & 4.1 & {\bf 1.3} & {\bf 1.3} & 4.9 & 4.8 & 3.8 & 3.8 \\
 $E_{\rm bond}$+$E_{\rm CI}$+LJ$_{\rm fit}$& 4.6 & 4.6 & 1.8 & 1.8 & 14.1 & 9.5 & 16.0 & 8.6 \\
 $E_{\rm bond}$+$E_{\rm PC}$+$E_{\rm pol}$+LJ$_{\rm fit}$ & 4.6 & 4.5 & 2.1 & 2.1 & 3.0 & 3.0 & {\bf 3.1} & {\bf 3.1} \\
 $E_{\rm bond}$+$E_{\rm MDCM}$+$E_{\rm pol}$+LJ$_{\rm fit}$ & 3.1 & 3.0 & {\bf 1.3} & {\bf 1.3} & {\bf 2.6} & {\bf 2.6} & 3.4 & 3.4 \\
  $E_{\rm bond}$+$E_{\rm MDCM}$+$E_{\rm pol}$+DE$_{\rm fit}$ & {\bf 2.9} & {\bf 2.9} & {\bf 1.3} & {\bf 1.3} & {\bf 2.6} & {\bf 2.6} & 3.2 & 3.2 \\
 $E_{\rm bond}$+$E_{\rm CI}$+$E_{\rm pol}$+LJ$_{\rm fit}$ & 4.0 & 4.0 & 1.8 & 1.8 & 9.8 & 9.6 & 9.6 & 8.2 \\
 $E_{\rm bond}$+$E_{\rm CI}$+$E_{\rm pol}$+DE$_{\rm fit}$ & 3.6 & 3.6 & 1.8 & 1.8 & 4.5 & 4.5 & 3.4 & 3.4 \\
 \hline
 2-body potential & 4.3 & 3.8 & 0.6 & 0.5 & 16.7 & 4.1 & 24.9 & 5.4 \\
 Corrected 2-body potential & {\bf 2.4} & {\bf 2.4} & {\bf 0.4} & {\bf 0.4} & {\bf 6.4} & {\bf 2.0} & {\bf 2.1} & {\bf 2.0} \\
 \hline
 \end{tabular}}
 \caption{RMSE and standard deviation $\sigma_{\rm error}$ (kcal/mol)
   in the error distribution are compared as terms in the original
   CHARMM force field are gradually replaced by reference values
   $E_{X}$ from electronic structure calculations. $X$ can be bonded,
   point charge (PC), MDCM, Coulomb integral (CI) or polarization
   (pol). Each time a term is replaced either the Lennard-Jones (LJ)
   or double-exponential (DE) repulsive terms are refitted to the
   total interaction energy. Results are compared with a 2-body
   potential with and without a many-body polarization correction. The
   lowest RMSE or $\sigma_{\rm error}$ in each class is shown in
   bold. Energies are reported in kcal/mol throughout.}
\bigskip {*The force field (FF) energy including bonded terms is
  compared with total DFT cluster energies by subtracting the mean of
  each distribution.}
\label{tab:comp}
\end{table}

\noindent
{\bf Electrostatic Terms:} Three different electrostatic models were
considered: ESP-fitted point charges (PCs), ESP-fitted minimally
distributed charge models (MDCMs) and an exact Coulomb integral based
on eq. \ref{Eq:Eel}. Each of these models can replace the
corresponding electrostatic term in the CHARMM energy function, with
the need to fit new LJ parameters to the remainder of the interaction
energy for each model. As the interaction energy and not the total
cluster energy is used as a reference, this is equivalent to a model
with reference PBE0 bonded terms, one of the improved electrostatic
terms, but with polarization, repulsion and other effects incorporated
into the remaining LJ term.\\

\begin{figure}
    \centering
    \includegraphics[width=1.0\textwidth]{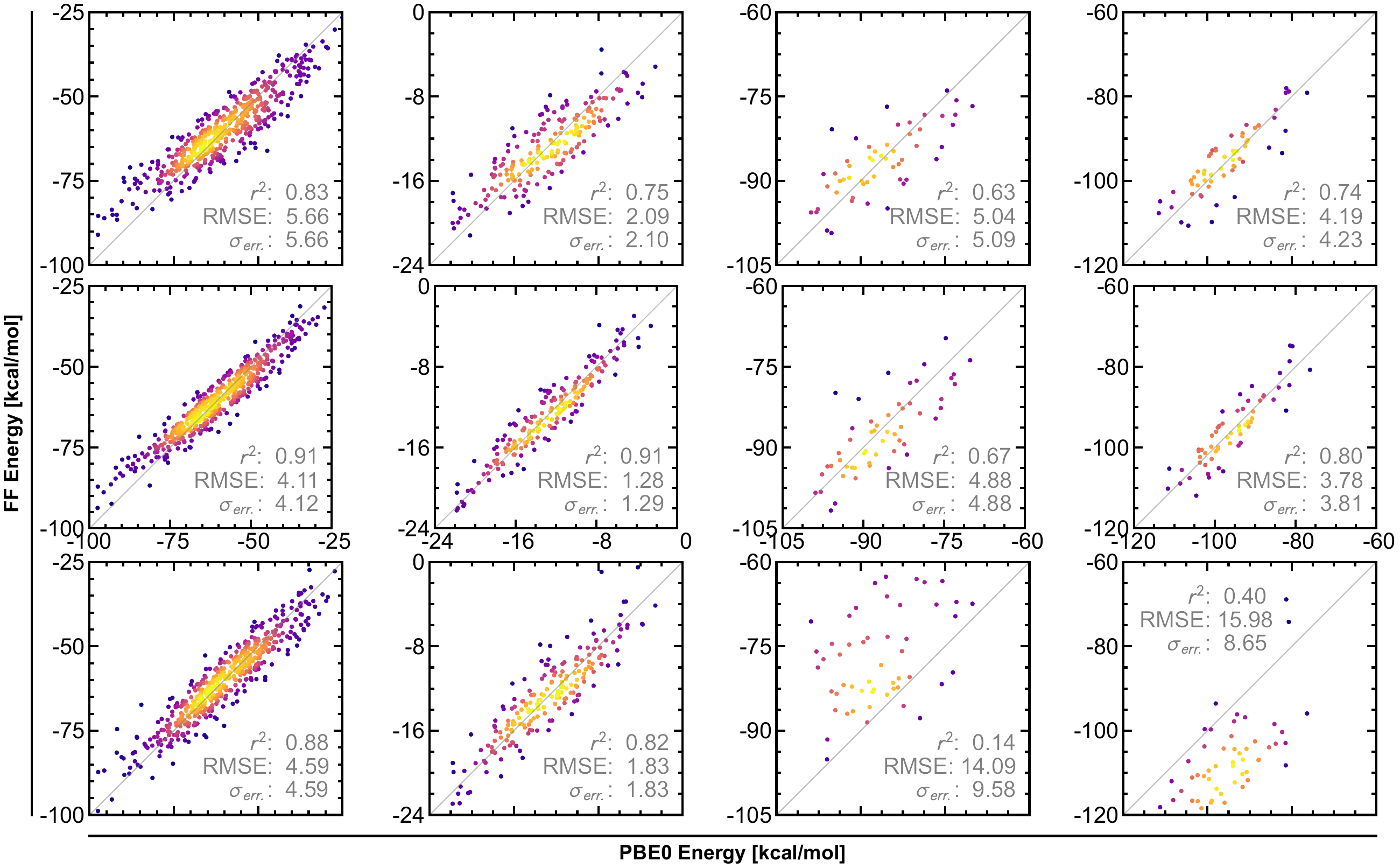}
    \caption{Comparison of force field cluster interaction energies
      ($y-$axis) using (top) PC, (middle) MDCM, and (bottom) Coulomb
      integral electrostatics for non-polarized monomers and fitted LJ
      parameters with interaction energies from reference PBE0
      calculations ($x-$axis) for water, dichloromethane, K$^+$ ions,
      and Cl$^-$ ions, from left to right. Color code: Yellow to
      orange to purple reflect decreasing density of points with
      yellow the maximum density.}
    \label{fig:cilj}
\end{figure}

\noindent
Results are summarized in Table \ref{tab:comp} and Figure
\ref{fig:cilj}. PC results use the standard CHARMM functional form for
nonbonded interactions, and as such represent the best possible model
for the unmodified functional form to describe the electronic
structure reference data. The fitting removes the systematic shift, as
visible in the very similar RMSE and $\sigma_{\rm error}$ values for
each cluster type.  Residual RMSEs of 4.2 to 5.8 kcal/mol for
H$_{2}$O, K$^{+}$ and Cl$^{-}$ are again reasonable, broadly
describing the relative energies of the different cluster geometries
correctly but still with an error distribution that leads to
significant changes in the ranking of cluster geometry energies. For
CH$_{2}$Cl$_{2}$ the RMSE is smaller at 2.1 kcal/mol due to the
decreased total interaction energies, but relative to the range in
energies the performance is similar.\\

\noindent
Introducing a more detailed MDCM electrostatic model to better
describe the ESP around each monomer yields improvement after also
refitting LJ parameters. This decreases the RMSE from 5.8 to 4.2
kcal/mol for H$_{2}$O, 2.1 to 1.3 kcal/mol for CH$_{2}$Cl$_{2}$, 5.0
to 4.9 kcal/mol for K$^{+}$ and 4.2 to 3.8 kcal/mol for Cl$^{-}$. In
the case of CH$_{2}$Cl$_{2}$ this is already the lowest RMSE that was
achieved through modification of the CHARMM force field.\\

\noindent
Finally, replacing the PC electrostatic interaction energy with an
exact Coulomb integral between frozen gas-phase monomers in their
cluster geometries was found to increase the RMSE relative to MDCM
results for H$_{2}$O and CH$_{2}$Cl$_{2}$, and to strongly increase
the RMSE and $\sigma_{\rm error}$ for K$^{+}$ and Cl$^{-}$ relative to
both MDCM and PC after refitting LJ parameters. This increase is
largely explained by the `penetration energy', that enhances the
electrostatic interaction due to reduced screening of nuclear charge
at short range. The more approximate PC and MDCM approaches do not
include a penetration energy correction and so short-range
electrostatic interaction energies are smaller and easier to balance
with short-range repulsion from the LJ term. In addition, the LJ
functional form was selected to work well with the PC approximation,
and not to balance the steeper electrostatic interaction profile of
the true Coulomb integral. This effect is discussed in more detail in
the van der Waals section below.\\

\begin{figure}
    \centering
    \includegraphics[width=1.0\textwidth]{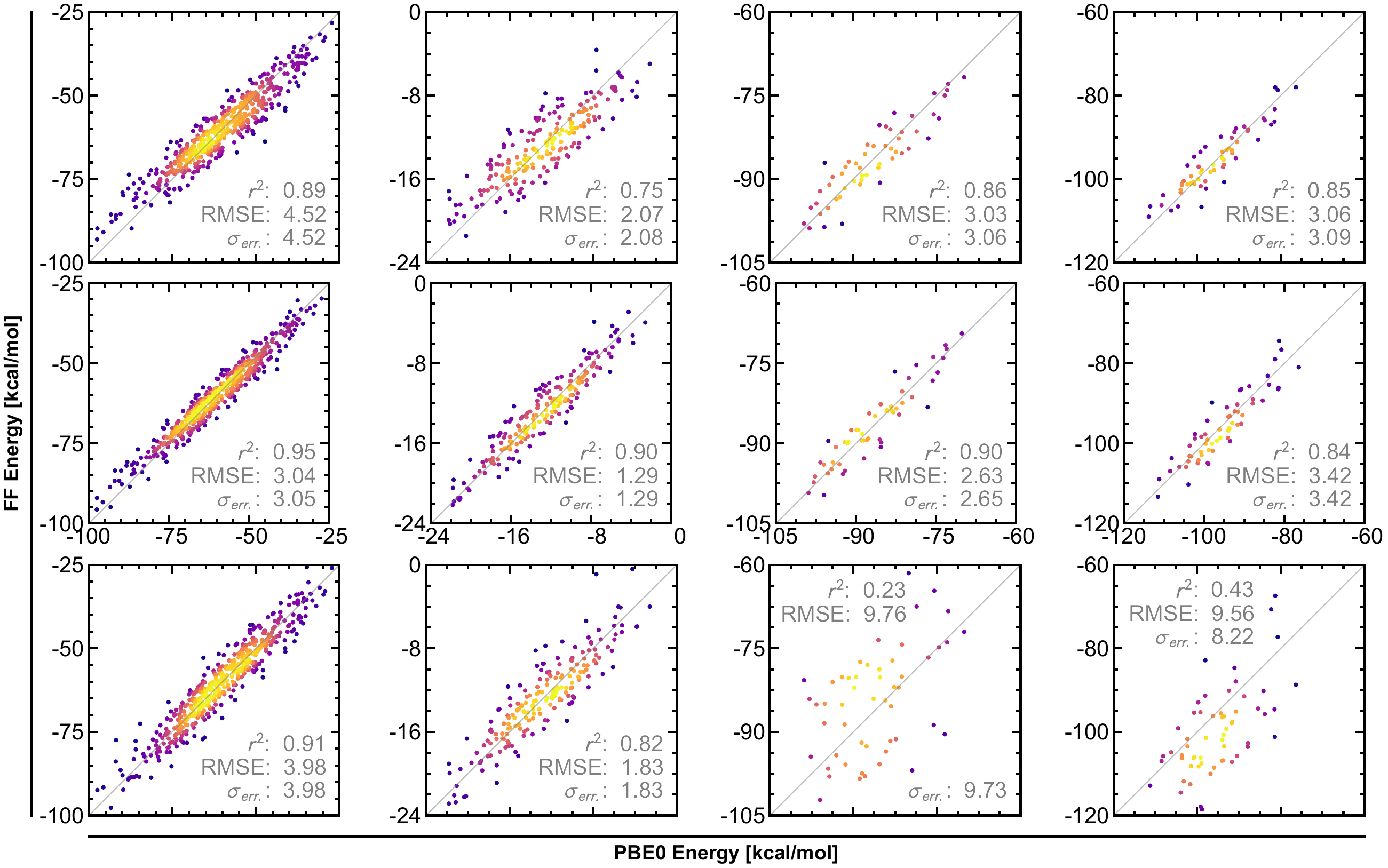}
    \caption{Comparison of PBE0 (x-axis) with FF (y-axis) interaction
      energies incorporating FMO polarization with PC (top), MDCM
      (middle), and Coulomb integral (bottom) electrostatics. Fitting
      results (left-right) for water, dichloromethane, potassium ions,
      and chloride ions. Color code: Yellow to orange to purple
      reflect decreasing density of points with yellow the maximum
      density.}
    \label{fig:pollj}
\end{figure}

\noindent
\label{S:polarization_results}
{\bf Electronic Polarization:} Adding a reference polarization term
(eq. \ref{Eq:Epol_FMO}) that depends only on the electric field from
each monomer in the supermolecule and refitting LJ parameters shows
how much of the residual error can be attributed to polarization
effects (see Table \ref{tab:comp} and Figure \ref{fig:pollj}). \\

\noindent
Results with PCs show a noticeable improvement in
RMSE (from 5.8 to 4.6 kcal/mol) after adding the polarization term for
H$_{2}$O, but without reaching the 4.2 kcal/mol accuracy of the
non-polarizable MDCM. This suggests that anisotropic electrostatics
are more important to capture than explicit polarization for
H$_{2}$O. In the case of CH$_{2}$Cl$_{2}$ there is no improvement,
suggesting polarization can be (implicitly) accounted for by using
only the LJ potential for pure liquid CH$_{2}$Cl$_{2}$. As might be
expected, an explicit polarization correction is
more important for the ionic systems and the polarization-corrected PC
models are now 1--2 kcal/mol more accurate than the non-polarizable
MDCM, suggesting explicit polarization is more important than
anisotropic electrostatics in describing surrounding solvent molecules
for charged systems.\\

\noindent
MDCM interaction energies are similarly improved, although again there
is no obvious benefit in adding explicit polarization in the case of
CH$_{2}$Cl$_{2}$, and Coulomb integral results are again degraded by
the lack of a suitable short-range repulsion term to balance the
penetration energy contribution.\\

\noindent
{\bf Van der Waals Contributions:} Results so far have used the LJ
term to describe dispersion and short-range Pauli repulsion. However,
by fitting to the total cluster energy the LJ term also contains
averaged, implicit corrections for the electrostatic penetration
energy, for the polarization energy (where not explicitly added), for
many-body effects and to correct for residual errors from the bonded
and nonbonded terms. This approach is not uncommon, whether by fitting
to reference energies or bulk-averaged properties from MD
simulations.\\

\begin{figure}[h!]
    \centering
    \includegraphics[width=0.75\textwidth]{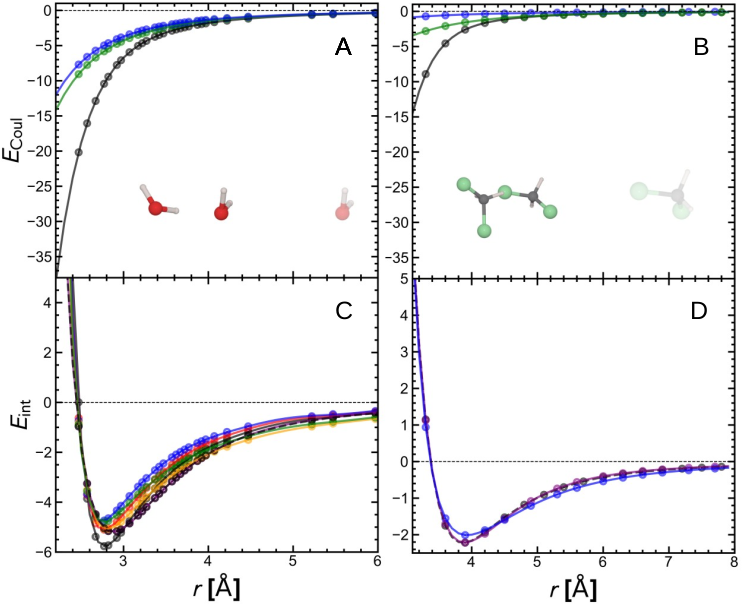}
    \caption{Electrostatic interaction energy (kcal/mol) along a
      minimum energy scan for dimers of water (panel A, $r$ is the
      O$\cdots$O separation) and dichloromethane (panel B, $r$ is the
      C$\cdots$C separation). Curves are calculated using PCs (blue),
      MDCM (green) without penetration energy corrections and
      reference Coulomb integral (black) data. Panel C: Water dimer
      interaction energy from reference PBE0 calculations (pink
      covered by grey, see below), PC/LJ (blue), MDCM/LJ (red),
      Coulomb integral/LJ (black), PC/DE (green), MDCM/DE (orange),
      and Coulomb integral/DE (grey). Panel D: For dichloromethane
      weaker electrostatics imply that the standard 12-6 LJ/CI
      potential (blue) reproduces the interaction better than in
      (panel C), whereas the DE potential/CI again models the
      interaction suitably. Note the different energy ranges
      ($y-$axis) in panels C and D.}
    \label{fig:dist-scan}
\end{figure}

\noindent
As already demonstrated for electrostatic interaction energies from
using the Coulomb integral, the 12-6 LJ functional form may no longer
be an optimal choice.\cite{kramer:2013,warshel:1970} Indeed,
polarizable force fields such as AMOEBA\cite{ponder:2010} use a 14-7
term in place of the Lennard-Jones 12-6 exponents. Figure
\ref{fig:dist-scan}A and B demonstrate the extent to which neglect of
a penetration energy correction for PC and MDCM models results in a
slower increase in Coulomb interaction energy at close range than in
the reference Coulomb integral. This in turn requires a shallower
repulsive potential to balance the short-range attraction and recover
the total interaction energy. Hence, the 12-6 LJ potential combined
with the Coulomb integral leads to larger errors for scanning the
energy for (H$_2$O)$_2$ along the O--O separation
in the H-bonded geometry (black curve in Figure
\ref{fig:dist-scan}C). PC and MDCM models that neglect penetration can
be better balanced in this case by the 12-6 potential. This leads to a
more accurate dimer energy.\\

\noindent
Combining the Coulomb integral with a double-exponential function
where the exponents are also fitted results in a significant
improvement for the H$_2$O dimer, as the Coulomb integral/DE total
energy (grey curve) becomes indistinguishable from the PBE0 dimer
reference energy (pink). In contrast, the DE functional form cannot
account for residual deficiencies in the description of the electric
field from PC and MDCM models. For these models the 12-6 functional
form already balances the electrostatic term reasonably well so there
is only marginal improvement in reproducing total dimer energies
(green and orange curves).\\

\noindent
Moving beyond the dimer, Table \ref{tab:comp} shows results for
combining MDCM and Coulomb integral electrostatics with a DE term in
the larger clusters. In contrast to the dimer scan, the full cluster
environments include a distribution of conformers as well as many-body
effects. The trends for H$_{2}$O remain nonetheless similar to the
dimer case, as replacing a LJ with a DE term in combination with
Coulomb integral electrostatics yields a modest improvement from 4.0
to 3.6 kcal/mol in the RMSE. For MDCM the RMSE only improves from 3.0
to 2.9 kcal/mol, again noticeably less than in the case of the Coulomb
integral. For K$^+$ and Cl$^-$ ions with Coulomb integrals in water
there is a more significant improvement than in pure H$_2$O: the
standard deviation of the error distribution, $\sigma_{\rm error}$, is
halved from 9.8 to 4.5 and 9.6 to 3.4 kcal/mol, respectively after
adding a DE term. Replacing the LJ with a DE term for MDCM in these
ionic systems results in much smaller improvement of maximally 0.2
kcal/mol, again showing that the 12-6 functional form is sufficient
when combined with MDCM but is not compatible with Coulomb integral
electrostatics. In contrast, the weaker electrostatics in
CH$_{2}$Cl$_{2}$ mean that there is no improvement in total
interaction energy from introducing a DE term with either Coulomb
integral or MDCM electrostatics, consistent with what was found for
the CH$_{2}$Cl$_{2}$ dimer, see Figure \ref{fig:dist-scan}D.\\

\noindent
In summary, the 12-6 LJ functional form appears to
be an acceptable choice for PC and MDCM models as there is limited
improvement in describing either dimer or total cluster interaction
energy upon refitting DE exponents. For the systems studied here, the
error in total energy with PC or MDCM electrostatics is instead
dominated by factors such as residual errors in the charge model in
describing the electric field, or non-polarization many-body effects
where polarization is accounted for. If the penetration energy is
explicitly included, as is the case in the exact Coulomb integral,
then the 12-6 functional form is no longer able to balance the steeper
short-range repulsion. This contribution to the total error becomes
particularly significant for polar or charged systems where the
electrostatic interaction energy dominates. Moreover, due to the
large absolute electrostatic energies involved at short range, even a
more flexible DE functional form combined with an exact Coulomb
integral can incur larger errors across a distribution of conformers
than simpler charge models with approximate descriptions of the
electric field that are damped at short range by neglecting
penetration.\\

\subsection{Correcting the 2-body Potential}
{\bf FMO Reference Data:} FMO-embedding is next used to provide a
many-body correction that depends only on the electric field of
surrounding monomers and neglects many-body Pauli contributions. As
such, the approximations inherent to FMO represent the best-possible
model for the polarization-corrected 2-body energy that will be
described here. As a similar polarization correction was added to the
CHARMM potential in section \ref{S:polarization_results} it is
important to establish the accuracy of the FMO-embedded dimer
approximation that is used to generate the corrective terms. Figure
\ref{fig:si_fmo_vs_supermol} compares FMO results with reference DFT
supermolecule calculations which yield residual $\sigma_{\rm error}$
values of $\sim 2$ kcal/mol (2\% of the total cluster interaction
energy). For solvated Cl$^{-}$ anions the RMSE is slightly larger as
there is a systematic shift in the FMO interaction energies of 2--3
kcal/mol. These errors remain small in comparison to the range of
interaction energies (ca. 80 kcal/mol in the polar systems), meaning
FMO represents a suitable choice for improving the PES by describing
the response to an embedding electric field.\\

\noindent
While further refinement of the FMO interaction energies would be
possible by moving beyond the embedded dimers approximation to also
include trimers, this would require any empirical function that models
it to also become 3-body in nature. This significantly increases the
training difficulty and the subsequent computational cost of
evaluating the interactions.\\

\begin{figure}
    \centering
   \includegraphics[width=0.8\textwidth]{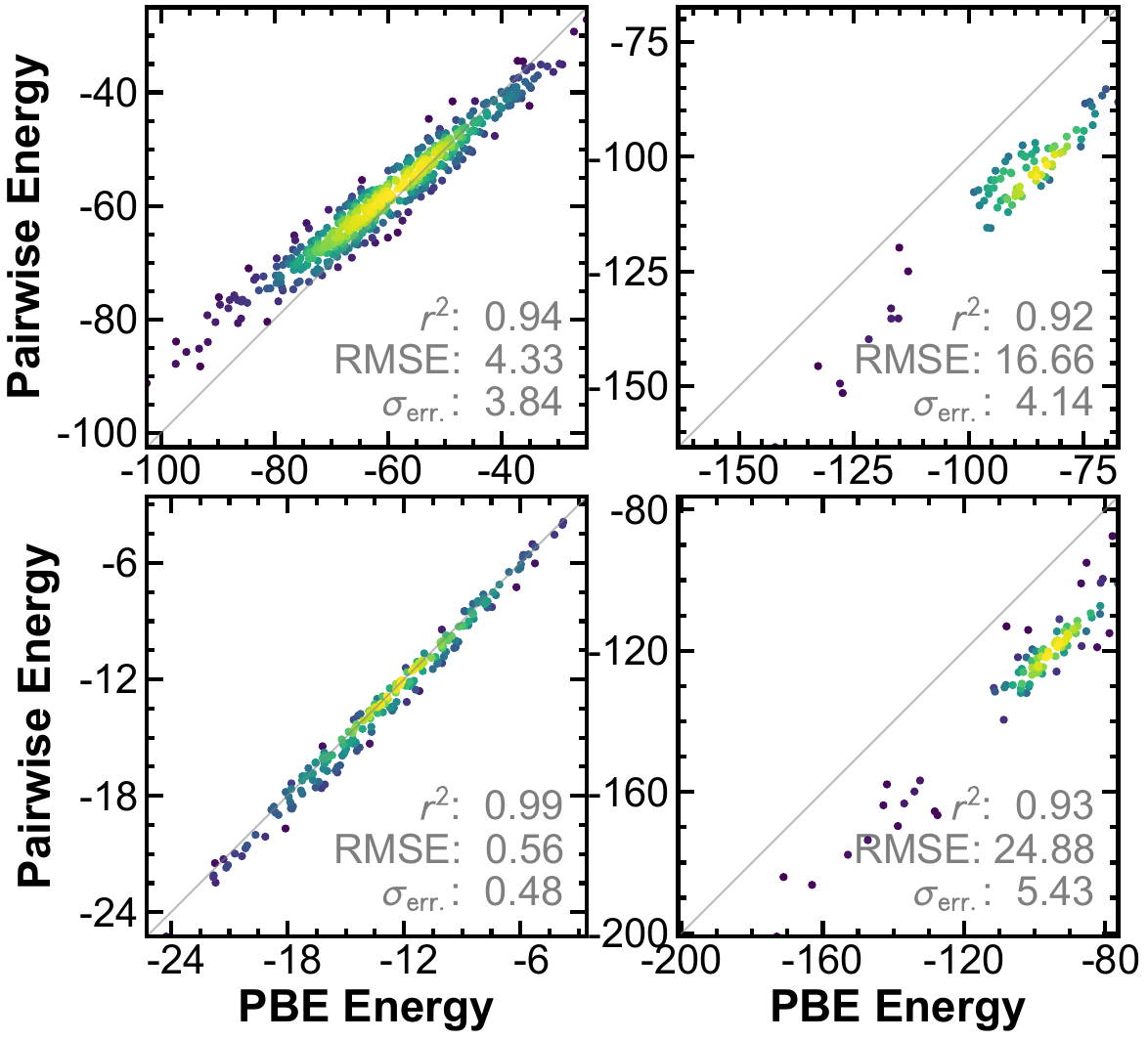}
    \caption{Comparison of PBE0/aug-cc-pVDZ supermolecule interaction
      energies and a 2-body approximation (sum over dimer energies)
      for pure water clusters (top-left), solvated K$^{+}$ ions
      (top-right), CH$_{2}$Cl$_{2}$ clusters (bottom-left) and
      solvated Cl$^{-}$ ions (bottom-right). Color code: Yellow to
      green to purple reflect decreasing density of points with yellow
      the maximum density.}
    \label{fig:pairwise_vs_supermol}
\end{figure}

\noindent
{\bf 2-body Approximation:} The `2-body' approximation, see
eq. \ref{Eq:2body}, assumes that the total energy of a system of
interacting monomers can be broken down into constituent dimers. Each
dimer interaction energy is evaluated by removing all other monomers,
and the total interaction energy is simply the sum of the dimer
interaction energies. Such an approximation can be quantitatively
tested by comparing the supermolecule interaction energy with the
corresponding pairwise dimer interaction sum for each cluster, see
Figure \ref{fig:pairwise_vs_supermol}.\\

\noindent
The results quantitatively show the limitations of pairwise-additive
force fields if many-body effects are not accounted for. The standard
deviation of the error distribution $\sigma_{\rm error}$ increases by
2 kcal/mol or more for H$_{2}$O, K$^+$ and Cl$^-$ relative to FMO
calculations, in line with previous results.\cite{stoneXantheas:2023}
For CH$_{2}$Cl$_{2}$ the difference is substantially smaller (ca. 0.1
kcal/mol), suggesting that many-body effects are not significant and a
2-body potential will suffice. Tests of this nature may also be
  useful for future parametrizations to determine whether or not
  two-body potentials are sufficiently accurate for particular systems
  investigated.\\

\noindent
In order to overcome the 2-body limit, force fields without explicit
many-body terms such as polarization may encapsulate this missing
many-body contribution in some averaged fashion. However, in so doing
the models will inevitably become environment-dependent, harming
transferability. The many-body correction introduced next is intended
to correct a 2-body ML surface, avoiding permutational issues in
training ML models.\\

\noindent
{\bf Corrected 2-body Potential:} Finally, the approximate 2-body
interaction is corrected by adding the FMO many-body polarization term
described in eq. \ref{Eq:mb_pol_corr}. This accounts for neglected
many-body effects on the total cluster interaction energy.  \\

\begin{figure}[!ht]
    \centering
    \includegraphics[width=0.8\textwidth]{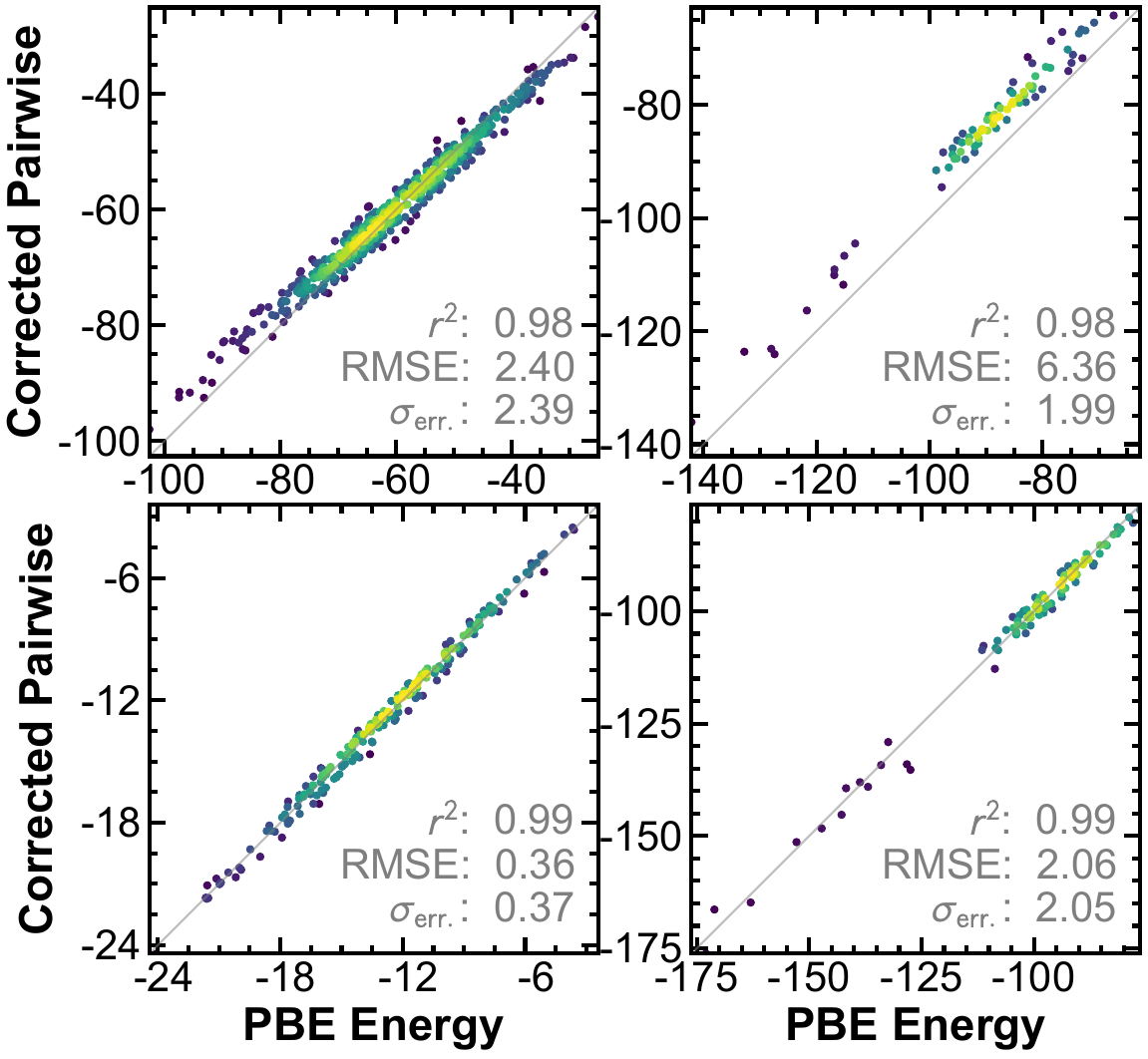}
    \caption{Comparison of 2-body description of total cluster
      interaction energy with an FMO many-body polarization correction
      and PBE0/aug-cc-pVDZ supermolecule cluster interaction
      energy. Results for water clusters (top-left), CH$_{2}$Cl$_{2}$
      (bottom-left), solvated K$^{+}$ (top-right), and Cl$^{-}$ ions
      (bottom-right) are shown. Color code: Yellow to green to purple
      reflect decreasing density of points with yellow the maximum
      density.}
    \label{fig:corrected_2body}
\end{figure}

\noindent
Figure \ref{fig:corrected_2body} shows the total cluster interaction
energy after adding the FMO many-body correction to the 2-body
approximation in Figure \ref{fig:pairwise_vs_supermol}. Again, it is
apparent that the many-body correction is not necessary for
CH$_{2}$Cl$_{2}$. The RMSE changes only by 0.2 kcal/mol relative to
Figure \ref{fig:pairwise_vs_supermol}. Figure
\ref{fig:si_pol_vs_2body} shows that this is due to the rather small
contribution of the polarization energy term to the total interaction
energy, spanning just 2 kcal/mol for all CH$_2$Cl$_2$ clusters. For
H$_2$O, the corrected dimer surface recovers the accuracy of an FMO
calculation (Figure \ref{fig:si_fmo_vs_supermol}) with a residual RMSE
of 2.4 kcal/mol when compared to a full supermolecule calculation, and
for the solvated ions the accuracy of the full FMO calculation is
again recovered. However, there is a small systematic shift: results
for K$^{+}$ reach an RMSE of 6.4 kcal/mol but with a narrow error
distribution ($\sigma_{\rm error}=2.0$), whereas results for Cl$^{-}$
are now better described than in the original FMO calculation as the
preexisting systematic shift is canceled out, with an RMSE of 2.1
kcal/mol.\\

\noindent
Overall, the data demonstrate that fitting directly to the dimer PES
and adding a subsequent FMO polarization correction depending only on
the local `embedding' electric field of the surrounding monomers is a
tractable approach to reach the accuracy of a full FMO calculation,
which in turn offers a good approximation to full supermolecule
interaction energies.\\

\subsection{Error Distributions}
While LJ and DE parameters fitted so far use the PES as reference
data, a common alternative is to fit primarily to bulk properties from
experiment and compare with results from molecular dynamics
simulations. An advantage of using experimental bulk data is firstly
that issues with choosing a reliable level of theory to generate the
PES are avoided. Secondly, such an approach implicitly focuses on
features of the PES that are most relevant to the set of bulk
properties in question. Bulk properties are often a function of an
energy distribution, and thus are not sensitive to achieving accurate
relative energies of conformational substates. For example, the
vaporization enthalpy is a function of the mean energy $\langle E_{\rm
  liq} \rangle$ of the sampled distribution of $N$ monomers $\Delta
H_{\rm vap} = \langle E_{\rm gas} \rangle - \langle E_{\rm liq}
\rangle / N + RT$.\\

\begin{figure}
    \centering
    \includegraphics[width=0.8\textwidth]{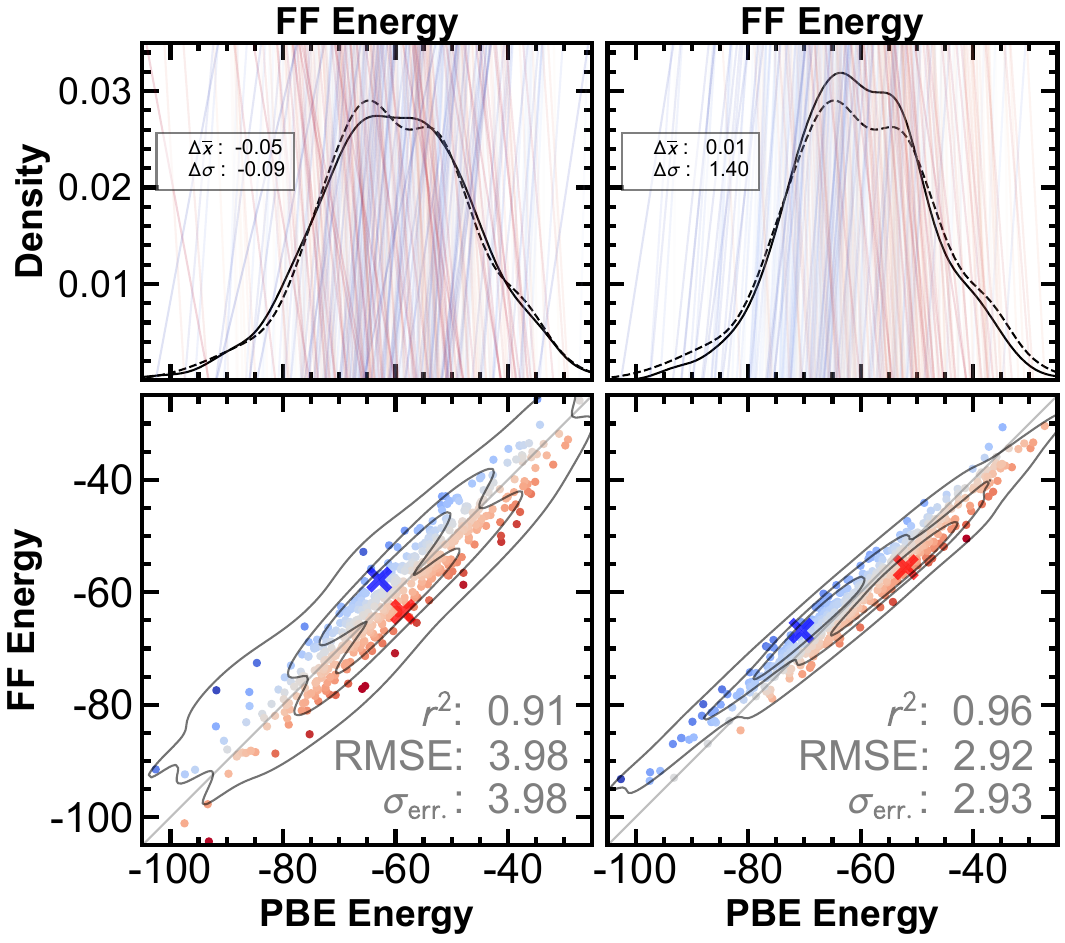}
    \caption{Fitted force field energies (FF Energy) using
      $E_\mathrm{CI+pol+LJ}$ (left) and $E_\mathrm{MDCM+pol+DE}$
      (right) electrostatics for H$_2$O. Top row: Kernel density
      estimates of the energy distributions for the reference (dashed)
      and the fitted model energies (solid). The vertical lines
      indicate positive (red) and negative (blue) differences between
      reference and model energies. The bottom $x-$axis are the PBE0
      reference energies and the top $x-$axis are the model
      energies. Hence, the slope of the lines directly reports on the
      magnitude of the difference. For vertical lines the reference
      and model energies are identical. The shift in mean and variance
      between the reference and fitted energy distributions is given
      as $\Delta \bar{x}$ and $\Delta \sigma$. Bottom row: Kernel
      density estimates for the positive (red) and negative (blue)
      differences between reference and model energies. The colored
      crosses give the center of gravity for the positive (red) and
      negative (blue) differences. RMSE, $r^2$ and $\sigma_{\rm err}$
      are provided.}
    \label{fig:error-distribs}
\end{figure}

\noindent
It is therefore of interest to investigate the extent to which a
faithful representation of an energy distribution coincides with
accurate descriptions of the individual energies contained in the
conformational ensemble. Figure \ref{fig:error-distribs} shows that
two electrostatic models, the Coulomb integral with polarization and
LJ model `$E_{\rm CI+pol+LJ}$' (left) with RMSE 4.0 kcal/mol and the
MDCM with polarization and DE model `$E_\mathrm{MDCM+pol+DE}$' (right)
with RMSE 2.9 kcal/mol, yield similar energy distributions. The
Coulomb model performs slightly better at describing the distribution,
in terms of closely matching both the mean and standard deviation of
the reference PBE0 energy distribution for the set of conformers
sampled. Representation of the individual conformer energies is
visibly worse for the Coulomb model, however, as evidenced by the
larger RMSE and $\sigma_{\rm error}$ values. The vertical lines in the
figures also map elements of the PBE0 reference data (bottom axis) to
corresponding FF data (top axis), so a vertical line at the far-right
of the distribution means that data in this region of the distribution
are highest in energy for both PBE0 and FF. A left-sloping line
starting at the far-right of the distribution means that data in this
part of the distribution were highest in energy for PBE0, but the same
structures were not ranked highest in energy by the FF. As such, this
figure shows that while both FF models create distributions that match
the PBE0 energy distribution closely, the crossed lines highlight that
individual cluster interaction energies are not accurately described
in the case of the Coulomb + polarization model.\\

\noindent
The two lower panels in Figure \ref{fig:error-distribs} report the
correlation between the reference energies ($x-$axis, PBE0) and the
models ($y-$axis, left for $E_{\rm CI+pol+LJ}$, and right for $E_{\rm
  MDCM+pol+DE}$). Again, blue and red symbols are for negative and
positive differences, respectively. The isocontours are from kernel
density estimates and the blue and red crosses denote the centers of
gravity for the respective distributions. For $E_{\rm CI+pol+LJ}$ in
H$_2$O it is found that the centers of gravity are close to the
average of the distribution, whereas for $E_{\rm MDCM+pol+DE}$ they
are displaced. In other words, the model with $E_{\rm CI+pol+LJ}$
describes under- and overestimations of the model relative to the
reference in a comparable fashion whereas using $E_{\rm MDCM+pol+DE}$
underestimates the reference energies for low-energy structures and
overestimates them for high-energy structures. This is also visible in
the top row when focusing on the red and blue vertical lines,
respectively. These observations act as a reminder that RMSE alone -
even when supplemented with the underlying distributions - does not
give a complete picture of model performance.\\

\noindent
Overall, these results highlight the need to balance fitting of
distributions to bulk reference data with fitting to the PES as
well. This ensures that both the distribution and also the individual
conformational energies that will determine sampling, thermodynamic
rates, equilibrium positions and reaction pathways of conformational
transitions are well described.\\

\section{Discussion and Conclusion}
In the current contribution the different terms of an empirical force
field, in this case CHARMM, that describe the PES were first replaced
systematically by quantities from electronic structure theory. Polar
H$_2$O was found to benefit most from using anisotropic MDCM
electrostatics in place of a PC model (1.6 kcal/mol improvement),
although an additional polarization term was similarly important (1.2
kcal/mol improvement). The charged K$^+$ and Cl$^-$ systems benefited
most from including a polarization term with anisotropy in the
corresponding water model of lesser concern for K$^+$ in particular
(2.0 kcal/mol vs 0.2 kcal/mol). This suggests that for K$^+$ a more
detailed description of the gas-phase H$_2$O monomer electric field
($1^{st}$ order term) is of lower importance to describing interaction
energy when the underlying density will be strongly perturbed and the
$2^{nd}$ order term is large. For CH$_2$Cl$_2$, anisotropic
electrostatics were most effective in reducing RMSE, likely due to
improved description of the sigma hole on Cl atoms, while explicit
polarization was of little benefit as the relatively small
polarization energies involved were absorbed implicitly into the LJ
term. The results for the four systems considered confirm that no one
term is of universal importance and results are system-specific, see
Table \ref{tab:comp}. Although bonded terms were well described and
contributed little to the total error in the systems studied here, the
small size and low complexity of the molecules (which all lack
torsional degrees of freedom, for example) mean that further work for
larger systems will be required to determine the generality of this
result.\\

\noindent
The 12-6 LJ functional form was found to be well suited for the PC and
MDCM electrostatic models that neglect penetration effects. Once such
effects are included, however, particular care must be taken to
balance the much larger electrostatic energies at close range and a
different functional form is necessary. A DE potential can be
particularly helpful here as it adds flexibility in describing the
important repulsive region.\cite{DE_2023} Improvement in PC and MDCM
results upon replacing LJ with DE is smaller, as in this case the
error in the interaction energy is dominated by effects such as a lack
of coupling between DE and monomer conformation or residual errors in
describing the monomer electric field. \\

\noindent
FMO embedding yields corrective terms, both to add explicit
polarization to non-polarizable force fields and to provide a
many-body correction to a purely 2-body potential. This will allow
generation of reference data for fitting empirical or ML energy
functions in the future. The terms derived are conceptually close to a
polarization response comprising the change in the Coulomb integral
between monomers after FMO embedding in the cluster electric field,
and the destabilization energy of the monomers due to distortion of
their electron distribution. The use of FMO embedding separates out
the response to an electric field, which will be broadly similar
whatever chemical species generates that field and is thus easily
parametrized, from Pauli effects that cannot be described in the
absence of explicit electrons without resorting to chemical atom-types
and the permutational difficulties that then arise in parametrizing
many-body terms for heterogeneous systems. The restriction of monomer
wave functions to use only their own monomer basis functions during
FMO calculations eliminates basis set superposition effects that arise
from using the full system basis for each monomer, and individual
monomers in the supermolecular system remain well defined, making this
a useful alternative to similar approaches from existing energy
decomposition analysis (EDA) methods.\cite{SAPT,morokuma:1976} \\

\noindent
PESs based on different approximate models may yield comparable energy
distributions for a given set of conformers. Fitting to data that are
primarily sensitive to distributions, such as averaged bulk properties
from MD simulations, without also fitting to the underlying PES risks
achieving accurate bulk properties without accurate underlying
energies of conformational substates. This has implications for
subsequent conformational sampling, such as the barrier heights
between different conformational substates. Indeed, recent models that
aimed at encapsulating both bulk properties and to improve description
of the underlying PES performed better over a range of temperatures
and pressures than models that described the PES more approximately
but a subset of bulk properties
well.\cite{Paesani:2023,wang:2013,xantheasPCCP:2023}\\

\noindent
The most accurate description of the reference PES was achieved using
a many-body-corrected 2-body potential. This is perhaps not
surprising, as the 2-body potential used was evaluated at the same
PBE0 level as the reference data, and the FMO correction devised
transformed the 2-body potential to an approximate FMO interaction
energy. The results demonstrate the promise of an approach based on an
accurate 2-body potential, to which a many-body correction is then
added. ML has proven effective in learning dimer
surfaces\cite{paesani:2013,paesani2022,Paesani:2023} where the
training data required is limited. FMO reference data offers a
corrective term that is a function only of the embedding electric
field of surrounding monomers. The response term could then be
approximated using an existing polarization term,\cite{Thole1981} or
again using ML techniques with electric field data as the input.\\

\noindent
In conclusion, the present results highlight the importance of
improving the description of electrostatics in widely used force
fields, and of adding explicit polarization terms in polar systems in
particular. Smaller improvements were observed upon correcting bonded
terms or introducing additional flexibility into the vdW functional
form, unless required to balance an explicit penetration energy
contribution. Further improvement is obtained by moving away from this
fixed functional form, however, to a polarization-corrected dimer
surface, using FMO embedding as reference data for the corrective
term. This approach is made feasible by advances in ML representations
of dimer surfaces, combined with ML-derived MDCM charge models to
describe an embedding electric field and an empirical or ML
description of the dimer response to that field. This will offer
improved accuracy over a fixed functional form while remaining
tractable to train for heterogeneous systems.\\

\section*{Supporting Information Available}
The Supporting Information includes scatter plots that show
performance of widely used FF parameters vs PBE0 reference data for
nonbonded interactions, and similar plots of FMO interaction energies
vs. full PBE0 supermolecule results. Raw 2-body vs many-body
polarization energies are also shown along with an evaluation of the
impact of counterpoise corrections. Sample Molpro and GAMESS/FMO input
files are presented to aid reproducibility of generated data.\\

\section*{Data Availability} 
The reference data and codes that allow reproduction of the findings
of this study will be made available at
\url{https://github.com/MMunibas/FF-Fit}.

\section*{Acknowledgment}
This work was partially supported by the Swiss National Science
Foundation through grants 200021\_188724, 200020\_219779, the NCCR
MUST (to MM), and the University of Basel.\\

\bibliography{refs}

\clearpage

\renewcommand{\thetable}{S\arabic{table}}
\renewcommand{\thefigure}{S\arabic{figure}}
\renewcommand{\thesection}{S\arabic{section}}
\renewcommand{\d}{\text{d}}
\setcounter{figure}{0}  
\setcounter{section}{0}

\noindent
{\bf Supporting Information: Systematic Improvement of Empirical
  Energy Functions in the Era of Machine Learning}\\

\section{Improved Bonded Terms}

\begin{figure}
    \includegraphics[width=0.8\textwidth]{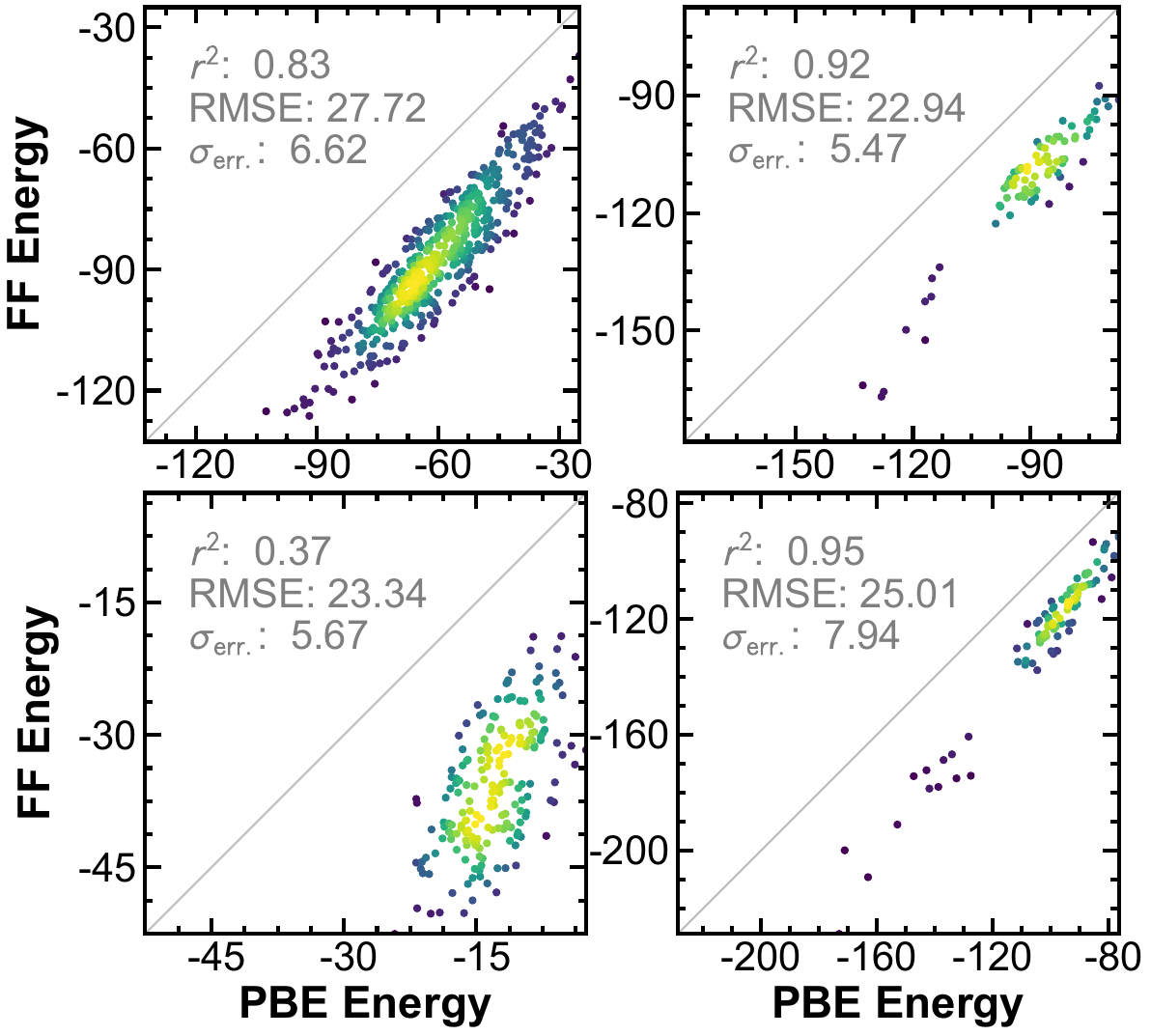}
    \caption{Comparison of cluster nonbonded energies (meaning errors
      from bonded interactions are removed) using standard CHARMM
      TIP3P and solvated ion parameters, CGENFF CH$_{2}$Cl$_2$
      parameters with reference PBE0 data. Results for water clusters
      (top-left), solvated K$^{+}$ ions (top-right), CH$_{2}$Cl$_{2}$
      (bottom-left) and solvated Cl$^{-}$ ions (bottom-right) are
      shown.}
    \label{fig:si_cgenff_vs_dft_nbonds}
\end{figure}

\newpage
\section{MDCM Models}

\begin{figure}
    \centering \includegraphics[width=0.8\textwidth]{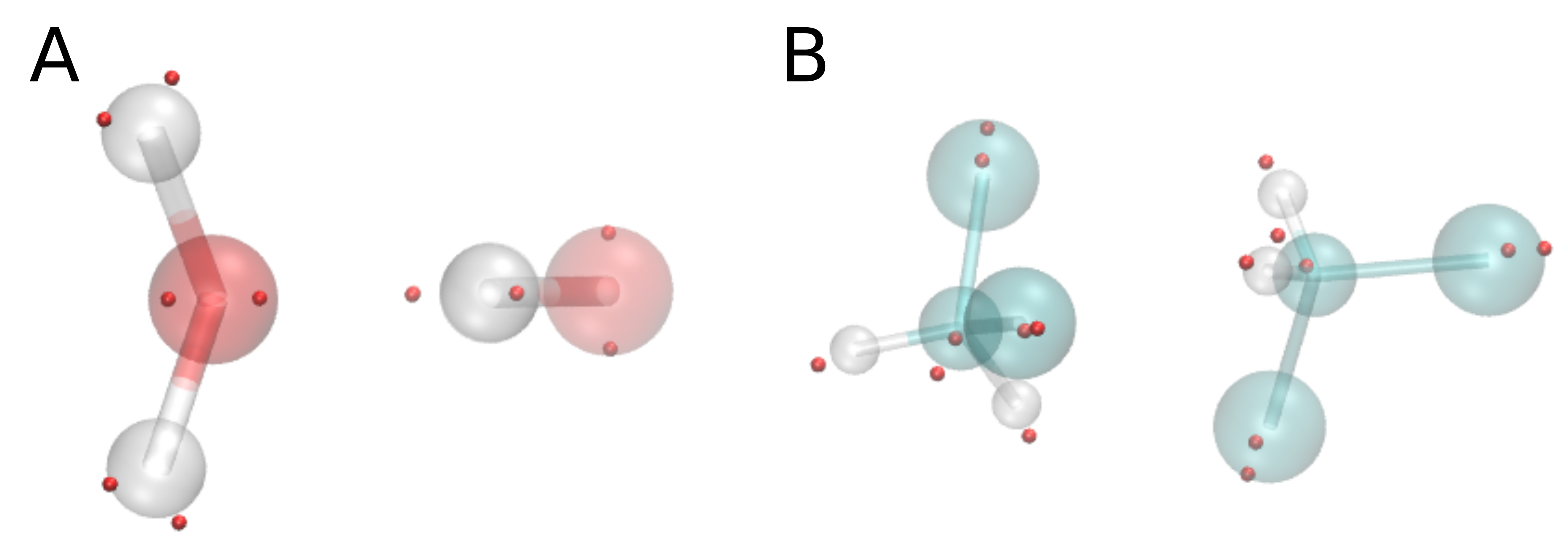}
    \caption{MDCM models for (A) water (six charges) and (B)
      dichloromethane (eight charges) optimized at the PBE/aug-cc-pVDZ
      level of theory. Red, white, and blue spheres are oxygen,
      hydrogen, and carbon atoms, respectively; the two large blue
      spheres are the chloride atoms in dichloromethane. The small red
      spheres are the positions of the MDCM charges.}
    \label{fig:si_mdcms}
\end{figure}

\newpage
\section{FMO vs PBE0 Supermolecule}

\begin{figure}
    \centering
    \includegraphics[width=0.8\textwidth]{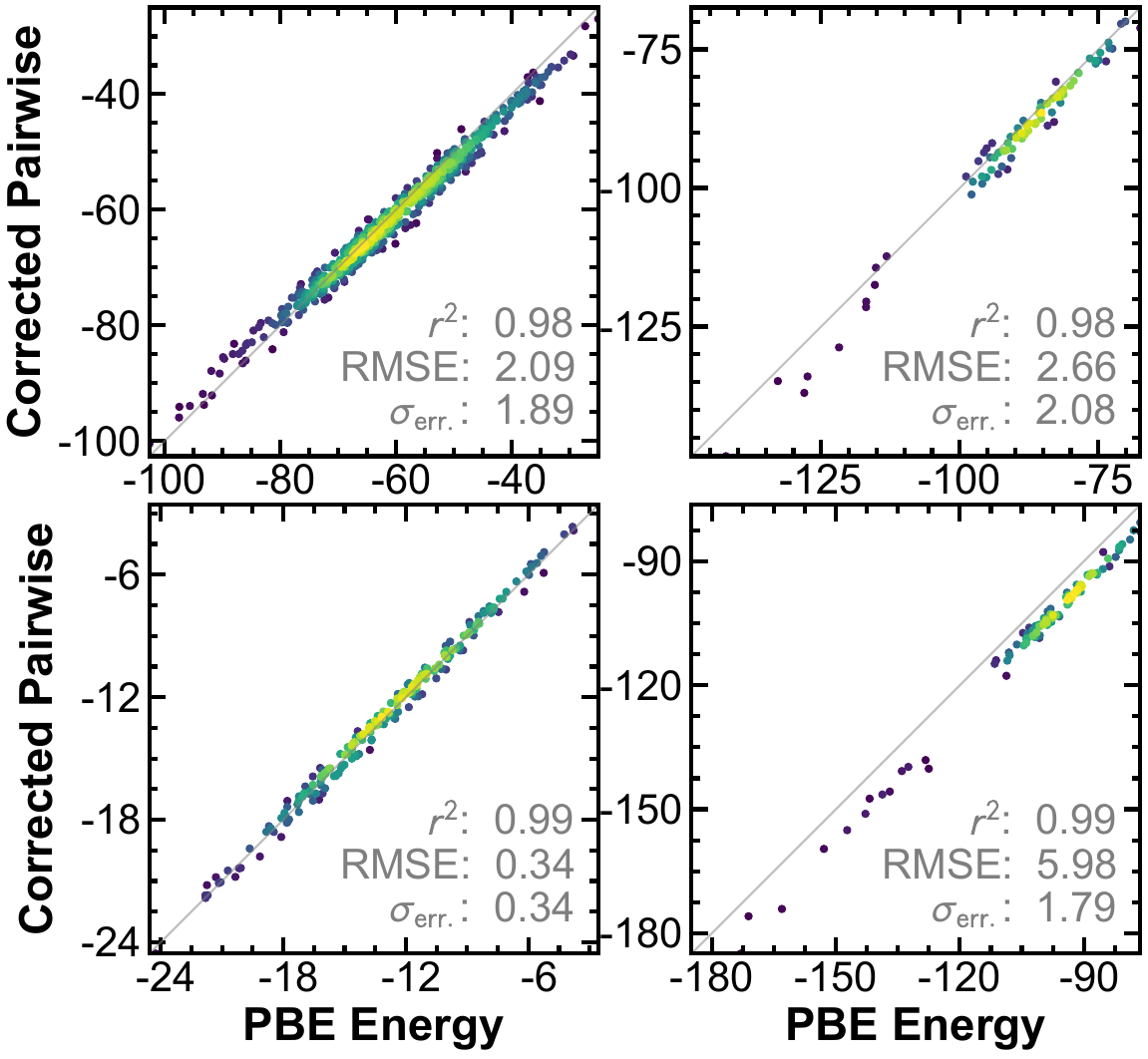}
    \caption{Comparison of PBE0/aug-cc-pVDZ supermolecule interaction
      energies and FMO interaction energies with restrained
      electrostatic potential point charge (RESPPC) embedding for pure
      water clusters (top-left), solvated K$^{+}$ ions (top-right),
      CH$_{2}$Cl$_{2}$ clusters (bottom-left) and solvated Cl$^-$ ions
      (bottom-right).}
    \label{fig:si_fmo_vs_supermol}
\end{figure}

\newpage
\section{Polarization Energy}

\begin{figure}[!htb]
    \centering
    \includegraphics[width=1.0\textwidth]{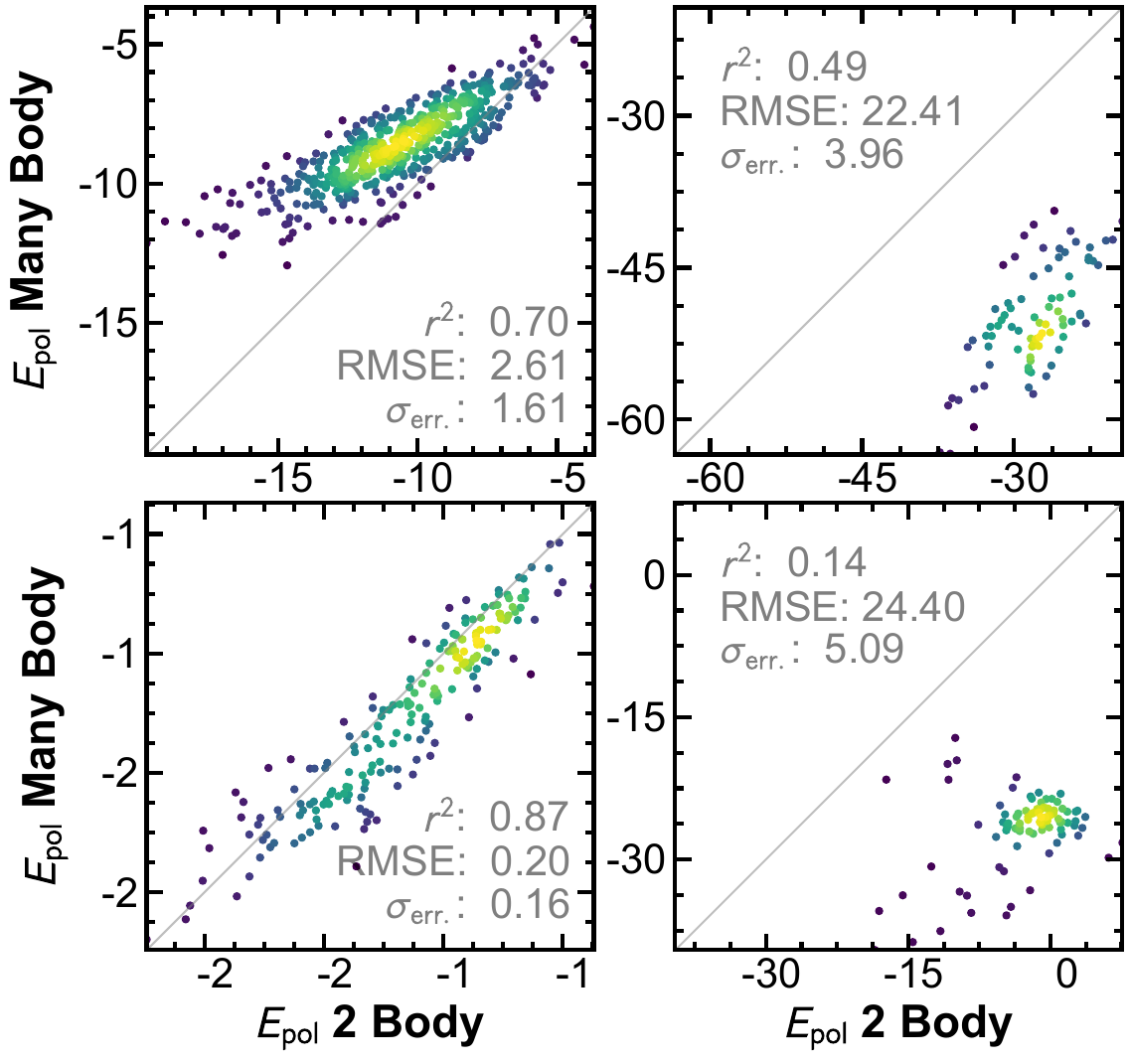}
    \caption{Comparison of sum over all 2-body polarization
      interactions with gas-phase monomer electrostatic interaction
      energies (Coulomb integrals) subtracted and many-body
      polarization energies between FMO monomers in the total cluster
      electric field, again with gas-phase monomer interaction
      energies subtracted. Results for water clusters (top left),
      solvated K$^{+}$ ions (top-right), CH$_{2}$Cl$_{2}$
      (bottom-left) and solvated Cl$^{-}$ ions (bottom-right) are
      shown.}
    \label{fig:si_pol_vs_2body}
\end{figure}

\newpage
\section{Basis Set Superposition Error}

\begin{figure}[!htb]
    \centering \includegraphics[width=1.0\textwidth]{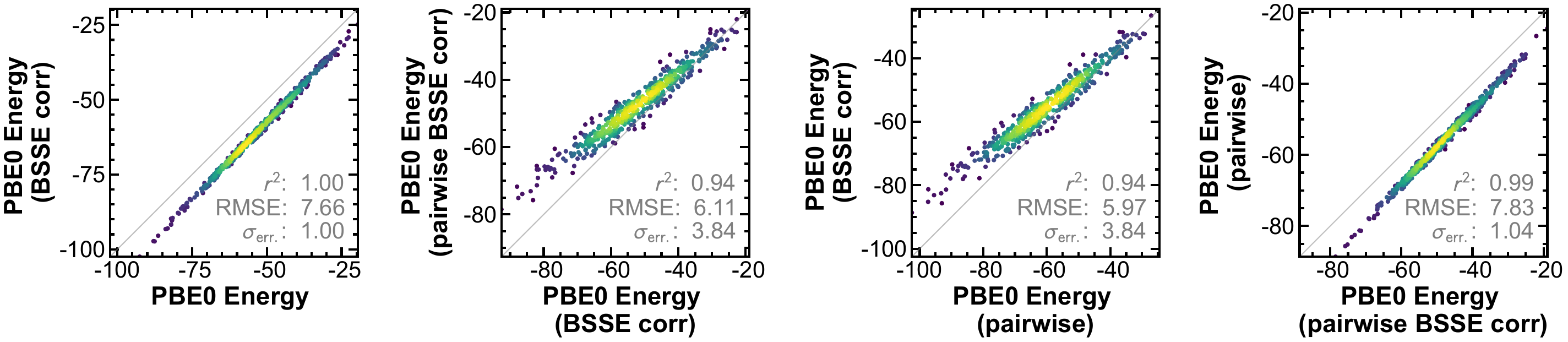}
    \caption{Comparison of total H$_{2}$O cluster interaction
      energies, left to right: without vs with counterpoise
      corrections, total counterpoise-corrected H$_{2}$O cluster
      interaction energy vs. total counterpoise-corrected pairwise
      interaction energy, total H$_{2}$O cluster interaction energy
      vs. total pairwise interaction energy without counterpoise (for
      comparison with panel 2), and pairwise H$_{2}$O cluster energy
      with and without counterpoise correction for each dimer
      interaction energy. }
    \label{fig:si_bsse}
\end{figure}

\newpage
\section{Manybody vs Pairwise Coulomb}

\begin{figure}[!htb]
    \centering \includegraphics[width=1.0\textwidth]{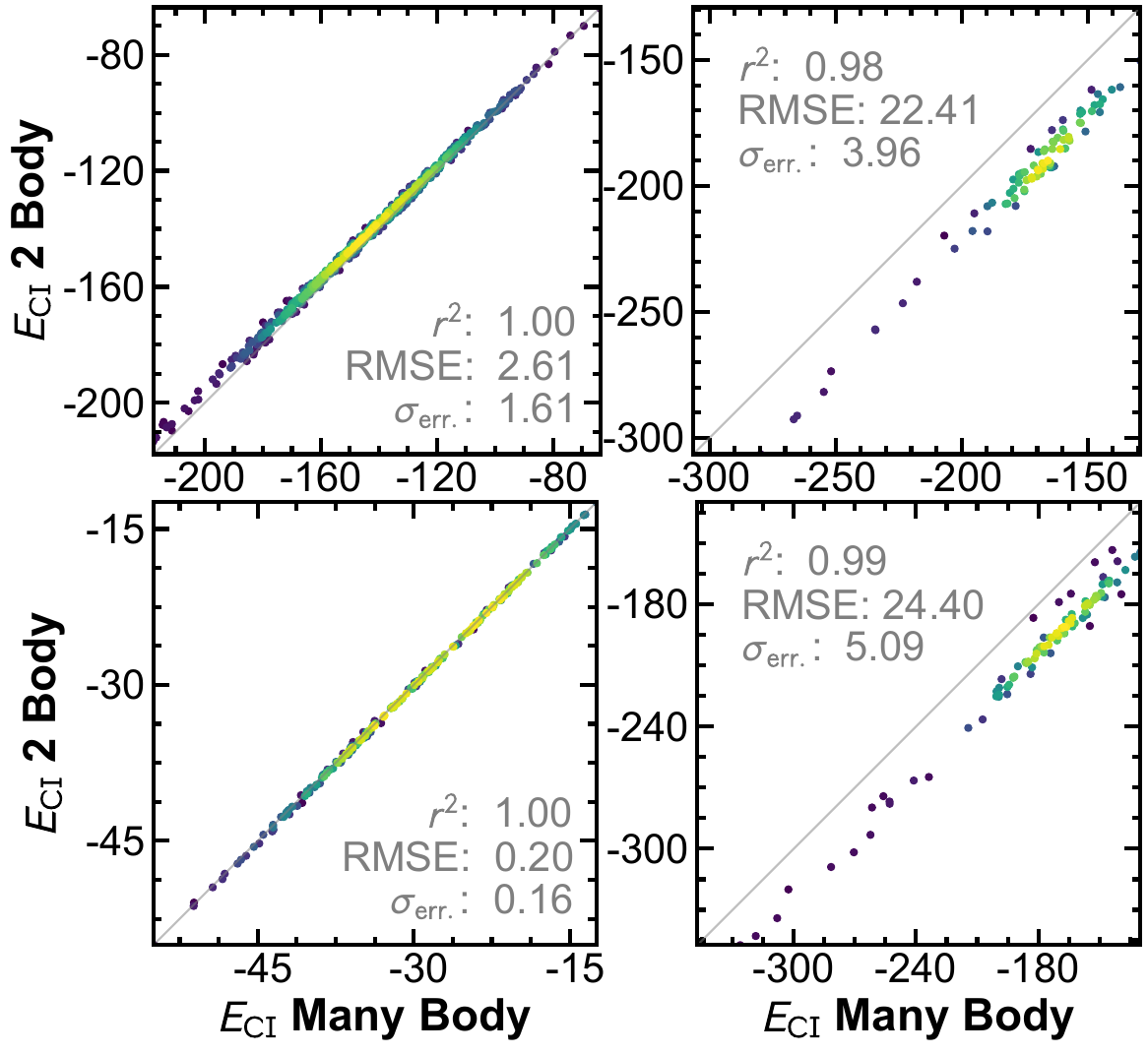}
    \caption{Comparison of sum over all 2-body electrostatic
      interactions (Coulomb integrals) between FMO monomers taken from
      dimer pairs (pure 2-body interactions) and total electrostatic
      interaction between FMO monomers in the total cluster electric
      field (including FMO embedding). Results for water clusters (top
      left), solvated K$^{+}$ ions (top-right), CH$_{2}$Cl$_{2}$
      (bottom-left) and solvated Cl$^-$ ions (bottom-right) are
      shown.}
    \label{fig:si_eel_pol_vs_2body}
\end{figure}

\newpage
\section{Sample Molpro Input}

\fontsize{12pt}{12pt}{
\begin{verbatim}
***,Molpro Input
gprint,basis,orbitals=50,civector
gthresh,printci=0.0,energy=1.d-8,orbital=1.d-8,grid=1.d-8
!gdirect
symmetry,nosym
orient,noorient

memory,150,m    ! memory: MEM_PER_CPU/8 * ~(0.90 to 0.95)
symmetry,nosym
orient,noorient
geometry={
angstrom;
***XYZ COORDINATES HERE***
}
charge=1
basis=avdz
gdirect;
{ks,pbe0}

edm=energy
name='40_19_POT_20_1.molden'
!  save xyz file
TEXT,$name
PUT,molden,$name,NEW
\end{alltt}
\end{verbatim}
}

\newpage
\section{Sample FMO Input}
\fontsize{12pt}{12pt}{
\begin{verbatim}
 $contrl nprint=7 ispher=1 dfttyp=pbe0 units=angs nosym=1
 maxit=100 qmttol=1e-5 itol=24 icut=12 $end
 $system mwords=100 memddi=100 $end
 $gddi ngroup=1 parout=.t. $end
 $scf  dirscf=.t. NPUNCH=0 diis=.f. soscf=.t. shift=.t. damp=.t.
 dirthr=1e-7 $end
 $fmoprp nprint=0 modorb=3 coroff=0 maxit=100 CONV=1e-5 $end
 $fmo nfrag=14 nlayer=1 NFRND=2 resppc=-1
FRGNAM(1)=
H2O01,H2O02,H2O03,H2O04,H2O05,H2O06,H2O07,H2O08,H2O09,H2O10,
H2O11,H2O12,H2O13,K14
ICHARG(1)=0,0,0,0,0,0,0,0,0,0,0,0,0,1
INDAT(1)=1,1,1,2,2,2,3,3,3,4,4,4,5,5,5,6,6,6,7,7,7,
8,8,8,9,9,9,10,10,10,11,11,11,12,12,12,13,13,13,14
 $end
 $data
14 H2O cluster
c1
*** BASIS SET HERE ***
 $end
 $fmoxyz
*** XYZ COORDINATES HERE ***
 $end
\end{verbatim}}

\newpage
\subsubsection{Basis set (corresponds to Molpro aug-cc-pVDZ builtin):}
\fontsize{10pt}{8pt}{
\begin{verbatim}
h-1 1
S   4
1         1.301000E+01           1.968500E-02
2         1.962000E+00           1.379770E-01
3         4.446000E-01           4.781480E-01
4         1.220000E-01           5.012400E-01
S   4
1  0.1301000000D+02  0.0000000000D+00
2  0.1962000000D+01  0.0000000000D+00
3  0.4446000000D+00  0.0000000000D+00
4  0.1220000000D+00  0.1000000000D+01
S   1
1         0.0297400              1.0000000
P   1
1         7.270000E-01           1.0000000
P   1
1         0.1410000              1.0000000

o-1 8
S   9
1         1.172000E+04           7.100000E-04
2         1.759000E+03           5.470000E-03
3         4.008000E+02           2.783700E-02
4         1.137000E+02           1.048000E-01
5         3.703000E+01           2.830620E-01
6         1.327000E+01           4.487190E-01
7         5.025000E+00           2.709520E-01
8         1.013000E+00           1.545800E-02
9         3.023000E-01          -2.585000E-03
S   9
1         1.172000E+04          -1.600000E-04
2         1.759000E+03          -1.263000E-03
3         4.008000E+02          -6.267000E-03
4         1.137000E+02          -2.571600E-02
5         3.703000E+01          -7.092400E-02
6         1.327000E+01          -1.654110E-01
7         5.025000E+00          -1.169550E-01
8         1.013000E+00           5.573680E-01
9         3.023000E-01           5.727590E-01
S    9
1  0.1172000000D+05  0.0000000000D+00
2  0.1759000000D+04  0.0000000000D+00
3  0.4008000000D+03  0.0000000000D+00
4  0.1137000000D+03  0.0000000000D+00
5  0.3703000000D+02  0.0000000000D+00
6  0.1327000000D+02  0.0000000000D+00
7  0.5025000000D+01  0.0000000000D+00
8  0.1013000000D+01  0.0000000000D+00
9  0.3023000000D+00  0.1000000000D+01
S   1
1         0.0789600              1.0000000
P   4
1         1.770000E+01           4.301800E-02
2         3.854000E+00           2.289130E-01
3         1.046000E+00           5.087280E-01
4         2.753000E-01           4.605310E-01
P   4
1  0.1770000000D+02  0.0000000000D+00
2  0.3854000000D+01  0.0000000000D+00
3  0.1046000000D+01  0.0000000000D+00
4  0.2753000000D+00  0.1000000000D+01
P   1
1         0.0685600              1.0000000
D   1
1         1.185000E+00           1.0000000
D   1
1         0.3320000              1.0000000

c-1 6
S   9
1         6.665000E+03           6.920000E-04
2         1.000000E+03           5.329000E-03
3         2.280000E+02           2.707700E-02
4         6.471000E+01           1.017180E-01
5         2.106000E+01           2.747400E-01
6         7.495000E+00           4.485640E-01
7         2.797000E+00           2.850740E-01
8         5.215000E-01           1.520400E-02
9         1.596000E-01          -3.191000E-03
S   9
1         6.665000E+03          -1.460000E-04
2         1.000000E+03          -1.154000E-03
3         2.280000E+02          -5.725000E-03
4         6.471000E+01          -2.331200E-02
5         2.106000E+01          -6.395500E-02
6         7.495000E+00          -1.499810E-01
7         2.797000E+00          -1.272620E-01
8         5.215000E-01           5.445290E-01
9         1.596000E-01           5.804960E-01
S    9
1         0.6665000000D+04  0.0000000000D+00
2         0.1000000000D+04  0.0000000000D+00
3         0.2280000000D+03  0.0000000000D+00
4         0.6471000000D+02  0.0000000000D+00
5         0.2106000000D+02  0.0000000000D+00
6         0.7495000000D+01  0.0000000000D+00
7         0.2797000000D+01  0.0000000000D+00
8         0.5215000000D+00  0.0000000000D+00
9         0.1596000000D+00  0.1000000000D+01
S   1
1         0.0469000              1.0000000
P   4
1         9.439000E+00           3.810900E-02
2         2.002000E+00           2.094800E-01
3         5.456000E-01           5.085570E-01
4         1.517000E-01           4.688420E-01
P    4
1         0.9439000000D+01  0.0000000000D+00
2         0.2002000000D+01  0.0000000000D+00
3         0.5456000000D+00  0.0000000000D+00
4         0.1517000000D+00  0.1000000000D+01
P   1
1         0.0404100              1.0000000
D   1
1         5.500000E-01           1.0000000
D   1
1         0.1510000              1.0000000

k-1 19
S   15
1  0.5233468000D+06  0.5846214466D-04
2  0.6784769000D+05  0.5459434170D-03
3  0.1410076000D+05  0.3164762417D-02
4  0.3747880000D+04  0.1409131076D-01
5  0.1165997000D+04  0.5137723924D-01
6  0.4054650000D+03  0.1506121150D+00
7  0.1532749000D+03  0.3253402485D+00
8  0.6201310000D+02  0.4035633083D+00
9  0.2634656000D+02  0.1903341454D+00
10  0.8550189000D+01  0.1375921051D-01
11  0.3440320000D+01 -0.1772691354D-02
12  0.8312492000D+00  0.5209943980D-03
13  0.3257631000D+00 -0.2730822086D-03
14  0.3636779000D-01  0.7594535801D-04
15  0.1731738000D-01 -0.4481963423D-04
S   15
1  0.5233468000D+06 -0.1675080339D-04
2  0.6784769000D+05 -0.1575410319D-03
3  0.1410076000D+05 -0.9065001834D-03
4  0.3747880000D+04 -0.4118730833D-02
5  0.1165997000D+04 -0.1510320306D-01
6  0.4054650000D+03 -0.4800820971D-01
7  0.1532749000D+03 -0.1159380235D+00
8  0.6201310000D+02 -0.2104140426D+00
9  0.2634656000D+02 -0.1020080206D+00
10  0.8550189000D+01  0.5140371040D+00
11  0.3440320000D+01  0.6018661218D+00
12  0.8312492000D+00  0.5077441027D-01
13  0.3257631000D+00 -0.1131300229D-01
14  0.3636779000D-01  0.2998840607D-02
15  0.1731738000D-01 -0.1755700355D-02
S   15
1  0.5233468000D+06  0.5529299171D-05
2  0.6784769000D+05  0.5179349224D-04
3  0.1410076000D+05  0.2998539551D-03
4  0.3747880000D+04  0.1352739797D-02
5  0.1165997000D+04  0.5022669247D-02
6  0.4054650000D+03  0.1585079762D-01
7  0.1532749000D+03  0.3958069407D-01
8  0.6201310000D+02  0.7226938917D-01
9  0.2634656000D+02  0.4123519382D-01
10  0.8550189000D+01 -0.2492509626D+00
11  0.3440320000D+01 -0.4463199331D+00
12  0.8312492000D+00  0.5989929102D+00
13  0.3257631000D+00  0.6174079075D+00
14  0.3636779000D-01  0.2295809656D-01
15  0.1731738000D-01 -0.1059319841D-01
S   15
1  0.5233468000D+06 -0.1076669576D-05
2  0.6784769000D+05 -0.1002579605D-04
3  0.1410076000D+05 -0.5854327696D-04
4  0.3747880000D+04 -0.2611318972D-03
5  0.1165997000D+04 -0.9842016127D-03
6  0.4054650000D+03 -0.3049098800D-02
7  0.1532749000D+03 -0.7828166920D-02
8  0.6201310000D+02 -0.1377129458D-01
9  0.2634656000D+02 -0.8916666491D-02
10  0.8550189000D+01  0.5276427924D-01
11  0.3440320000D+01  0.8958136475D-01
12  0.8312492000D+00 -0.1328289477D+00
13  0.3257631000D+00 -0.3100708780D+00
14  0.3636779000D-01  0.6576787412D+00
15  0.1731738000D-01  0.4475538239D+00
S   15
1  0.5233468000D+06  0.0000000000D+00
2  0.6784769000D+05  0.0000000000D+00
3  0.1410076000D+05  0.0000000000D+00
4  0.3747880000D+04  0.0000000000D+00
5  0.1165997000D+04  0.0000000000D+00
6  0.4054650000D+03  0.0000000000D+00
7  0.1532749000D+03  0.0000000000D+00
8  0.6201310000D+02  0.0000000000D+00
9  0.2634656000D+02  0.0000000000D+00
10  0.8550189000D+01  0.0000000000D+00
11  0.3440320000D+01  0.0000000000D+00
12  0.8312492000D+00  0.0000000000D+00
13  0.3257631000D+00  0.0000000000D+00
14  0.3636779000D-01  0.0000000000D+00
15  0.1731738000D-01  0.1000000000D+01
S   1
1  0.8250000000D-02  0.1000000000D+01
P   11
1  0.1036177000D+04  0.1781460518D-02
2  0.2341423000D+03  0.1561190454D-01
3  0.7317656000D+02  0.7644392222D-01
4  0.2674509000D+02  0.2354400684D+00
5  0.1053356000D+02  0.4309361253D+00
6  0.4296808000D+01  0.3715701080D+00
7  0.1626117000D+01  0.7333752132D-01
8  0.6527073000D+00 -0.4334011260D-02
9  0.2483930000D+00  0.1794580522D-02
10  0.4411967000D-01 -0.4305081252D-03
11  0.1585962000D-01  0.1913030556D-03
P   11
1  0.1036177000D+04 -0.5455521428D-03
2  0.2341423000D+03 -0.4780361251D-02
3  0.7317656000D+02 -0.2405980630D-01
4  0.2674509000D+02 -0.7617631993D-01
5  0.1053356000D+02 -0.1495530391D+00
6  0.4296808000D+01 -0.1200860314D+00
7  0.1626117000D+01  0.2381750623D+00
8  0.6527073000D+00  0.5550991453D+00
9  0.2483930000D+00  0.3566180933D+00
10  0.4411967000D-01  0.2228060583D-01
11  0.1585962000D-01 -0.6519991706D-02
P   11
1  0.1036177000D+04  0.7539442617D-04
2  0.2341423000D+03  0.6589882287D-03
3  0.7317656000D+02  0.3331941157D-02
4  0.2674509000D+02  0.1053260366D-01
5  0.1053356000D+02  0.2089780725D-01
6  0.4296808000D+01  0.1631180566D-01
7  0.1626117000D+01 -0.3688461280D-01
8  0.6527073000D+00 -0.8830043065D-01
9  0.2483930000D+00 -0.7923222750D-01
10  0.4411967000D-01  0.4194251456D+00
11  0.1585962000D-01  0.6694432324D+00
P   11
1  0.1036177000D+04  0.0000000000D+00
2  0.2341423000D+03  0.0000000000D+00
3  0.7317656000D+02  0.0000000000D+00
4  0.2674509000D+02  0.0000000000D+00
5  0.1053356000D+02  0.0000000000D+00
6  0.4296808000D+01  0.0000000000D+00
7  0.1626117000D+01  0.0000000000D+00
8  0.6527073000D+00  0.0000000000D+00
9  0.2483930000D+00  0.0000000000D+00
10  0.4411967000D-01  0.0000000000D+00
11  0.1585962000D-01  0.1000000000D+01
P   1
1  0.5700000000D-02  0.1000000000D+01
D   5
1  0.1799834000D+01  0.1882899149D-01
2  0.6122067000D+00  0.4320508047D-01
3  0.1366237000D+00  0.7650096541D-01
4  0.5649686000D-01  0.2132819036D+00
5  0.1585889000D-01  0.8410786198D+00
D   5
1  0.1799834000D+01  0.0000000000D+00
2  0.6122067000D+00  0.0000000000D+00
3  0.1366237000D+00  0.0000000000D+00
4  0.5649686000D-01  0.0000000000D+00
5  0.1585889000D-01  0.1000000000D+01
D   1
1  0.6340000000D-02  0.1000000000D+01

cl-1 17
S   12
1    127900.0000000              0.241153E-03
2     19170.0000000              0.187095E-02
3      4363.0000000              0.970827E-02
4      1236.0000000              0.393153E-01
5       403.6000000              0.125932E+00
6       145.7000000              0.299341E+00
7        56.8100000              0.421886E+00
8        23.2300000              0.237201E+00
9         6.6440000              0.191531E-01
10        2.5750000             -0.334792E-02
11        0.5371000              0.929883E-03
12        0.1938000             -0.396379E-03
S   12
1    127900.0000000             -0.678922E-04
2     19170.0000000             -0.521836E-03
3      4363.0000000             -0.276513E-02
4      1236.0000000             -0.111537E-01
5       403.6000000             -0.385919E-01
6       145.7000000             -0.994848E-01
7        56.8100000             -0.201392E+00
8        23.2300000             -0.130313E+00
9         6.6440000              0.509443E+00
10        2.5750000              0.610725E+00
11        0.5371000              0.421549E-01
12        0.1938000             -0.923427E-02
S   12
1    127900.0000000              0.204986E-04
2     19170.0000000              0.158298E-03
3      4363.0000000              0.833639E-03
4      1236.0000000              0.339880E-02
5       403.6000000              0.116738E-01
6       145.7000000              0.309622E-01
7        56.8100000              0.629533E-01
8        23.2300000              0.460257E-01
9         6.6440000             -0.219312E+00
10        2.5750000             -0.408773E+00
11        0.5371000              0.638465E+00
12        0.1938000              0.562362E+00
S   12
1    127900.0000000              0.000000E+00
2     19170.0000000              0.000000E+00
3      4363.0000000              0.000000E+00
4      1236.0000000              0.000000E+00
5       403.6000000              0.000000E+00
6       145.7000000              0.000000E+00
7        56.8100000              0.000000E+00
8        23.2300000              0.000000E+00
9         6.6440000              0.000000E+00
10        2.5750000              0.000000E+00
11        0.5371000              0.000000E+00
12        0.1938000              0.100000E+01
S   1
1         0.0608000              1.0000000
P   8
1       417.6000000              0.525982E-02
2        98.3300000              0.398332E-01
3        31.0400000              0.164655E+00
4        11.1900000              0.387322E+00
5         4.2490000              0.457072E+00
6         1.6240000              0.151636E+00
7         0.5322000              0.181615E-02
8         0.1620000              0.188296E-02
P   8
1       417.6000000             -0.143570E-02
2        98.3300000             -0.107796E-01
3        31.0400000             -0.470075E-01
4        11.1900000             -0.111030E+00
5         4.2490000             -0.153275E+00
6         1.6240000              0.894609E-01
7         0.5322000              0.579444E+00
8         0.1620000              0.483272E+00
P   8
1       417.6000000              0.000000E+00
2        98.3300000              0.000000E+00
3        31.0400000              0.000000E+00
4        11.1900000              0.000000E+00
5         4.2490000              0.000000E+00
6         1.6240000              0.000000E+00
7         0.5322000              0.000000E+00
8         0.1620000              1.000000E+00
P   1
1         0.0466000              1.0000000
D   1
1         0.6000000              1.0000000
D   1
1         0.1960000              1.0000000
\end{verbatim}}

\end{document}